\documentclass[12pt,a4paper]{article}%
\pdfoutput=1

\usepackage[T1]{fontenc} 
\usepackage[utf8]{inputenc}
\usepackage{amsmath}
\usepackage{amssymb}
\usepackage{amsthm}
\usepackage[ 
    ]{hyperref}
\usepackage{mathtools}
\usepackage{graphicx}
\usepackage{booktabs}
\usepackage{caption}
\usepackage{slashed}
\usepackage{listings}
\usepackage{slashed}
\usepackage{rotating}
\usepackage{subfig}
\usepackage{geometry}
\usepackage{multicol}
\usepackage[nottoc]{tocbibind}
\usepackage{multicol}
\usepackage{authblk}
\usepackage{cite}
\usepackage{appendix}
\usepackage{color}
\usepackage{epsfig}
\usepackage{hyperref}
\usepackage{enumerate}
\usepackage{amsfonts}
\usepackage{epstopdf}
\usepackage{latexsym}
\usepackage{arydshln}

\definecolor{myred}{rgb}{0.7, 0, 0}
\definecolor{myblue}{rgb}{0, 0, 0.7}
\definecolor{mygreen}{rgb}{0.04, 0.7, 0.5}

\hypersetup{colorlinks,citecolor=myred,linkcolor=myblue,urlcolor=myblue,linktocpage=true}

\newcommand{\be}{\begin{equation}}
\newcommand{\ee}{\end{equation}}
\newcommand{\bea}{\begin{eqnarray}}
\newcommand{\eea}{\end{eqnarray}}

\numberwithin{equation}{section}

\title{\Large \textbf{
Peccei-Quinn Phase Transition at LIGO}}

\author{
{Benedict von Harling$^{a}$,  Alex Pomarol$^{a,b}$, Oriol Pujol\`as$^{a}$\\ and Fabrizio Rompineve$^{c}$}\\
\normalsize\itshape $^a$ IFAE and BIST, Universitat Aut\`onoma de Barcelona, 08193~Bellaterra,~Barcelona\\
\normalsize\itshape $^b$ Dept.~de~F\'isica, Universitat Aut{\`o}noma de Barcelona, 08193~Bellaterra,~Barcelona\\
\normalsize\itshape $^c$ Institute of Cosmology, Dept.~of Physics and Astronomy, \\
\normalsize\itshape Tufts University, Medford,
\normalsize\itshape MA 02155, USA\\
}

\begin{document}

\maketitle

\begin{abstract}

\noindent The LIGO observatories can potentially detect stochastic gravitational waves   arising from phase transitions 
which happened in the early universe at  temperatures around $T\sim 10^{8}$ GeV.
This provides  an extraordinary    opportunity for discovering the
phase transition associated with the  breaking of the Peccei-Quinn symmetry,  required in QCD axion models.
Here we consider  the simplest Peccei-Quinn models and study under which conditions a strong first-order phase transition 
can occur,  analyzing its associated gravitational wave signal.
To be detectable at LIGO, we show that some  supercooling is needed, which can arise either
in Coleman-Weinberg-type symmetry breaking or in strongly-coupled models. 
We also  investigate  phase transitions that interestingly  proceed  by first breaking the electroweak symmetry at large scales before tunneling to the Peccei-Quinn breaking vacuum.
In this case, the associated gravitational wave signal is more likely to be probed at the proposed Einstein Telescope.
%

\noindent

\end{abstract}

\newpage

\section{Introduction}

The recent detection of gravitational waves (GWs) by LIGO \cite{Abbott:2016blz} represents the beginning of a new era in the exploration of the universe. In a few years LIGO-VIRGO has compiled a sizable catalogue of detected binary merger events \cite{LIGOScientific:2018mvr}, and the prospects to further increase the sensitivity and even to build more observatories look promising. 

In the zoo of candidates for GW signals there is one that stands out from the point of view of high-energy physics: the stochastic GW backgrounds originating from cosmological first-order phase transitions   in the early universe.
First-order phase transitions develop by the  formation of bubbles that expand, collide and percolate. 
The bubble wall collisions are violent events that occur everywhere in space at a given cosmological time,
leading  to  sizable stochastic signals 
that remain as a relic cosmological background analogous to the cosmic microwave background, but in GWs. 
Since GWs are a form of radiation, after their production the fraction of the energy density that they carry keeps constant in the radiation dominated epoch, thereby giving a relic background that can be detected now, no matter how early they were produced and how high the temperature of the universe was. 
The temperature of the phase transition is encoded directly into the power spectrum of the signal, mainly
in  the peak frequency  that scales as $f_{\rm peak}\propto T$. 
A first-order phase transition at $T\sim$ TeV  peaks in the  frequency sensitivity band of LISA, while GW observatories with higher frequency sensitivity bands can probe even higher energies \cite{Grojean:2006bp}. 

The main motivation for this work is that the LIGO frequency band corresponds to 
first-order phase transitions which could have happened when the early universe was at a temperature around $10^7-10^8$ GeV. This roughly coincides with the lowest possible energy scale where the Peccei-Quinn (PQ) symmetry $U(1)_{PQ}$ had to be broken in  QCD axion models which solve the strong CP problem of the SM~\cite{Peccei:1977hh, Peccei:1977ur}. In other words, the axion solution to the strong CP problem {\em predicts} a phase transition that can  occur  around this scale. 
Then, LIGO-VIRGO has the chance to discover this  PQ phase transition if  
it was of  first-order and "strong enough". 

The purpose of this work, then, is to browse through the simplest incarnations of the PQ mechanism and see in which cases a detectable first-order phase transition is obtained.  
We will focus on the minimal KSVZ~\cite{Kim:1979if,Shifman:1979if} and DFSZ~\cite{Dine:1981rt,Zhitnitsky:1980tq} models, as well as supersymmetric and strongly coupled versions of them.
As we will see, the most important requirement is that the models manage to give a strong enough (and long enough) transition.
We will show that this is the case for certain regions of the parameter space of the DFSZ model, and is more favorable, 
when the  PQ breaking is driven by the Coleman-Weinberg mechanism \cite{Coleman:1973jx}. 
Also, we will show that   strongly-coupled models of PQ breaking lead to long periods of supercooling  which end
with strong  GW signals  detectable  at LIGO.

Furthermore, we investigate the occurrence of a two-step PQ phase transition in DFSZ constructions, with an intermediate second-order    electroweak phase transition 
 at very high scales, before 
ending in the PQ broken minimum
with a  first-order phase transition. 
Crucially, we show that it is possible to obtain  a significant amount of "cooling" in these cases, 
 albeit much milder than in the aforementioned supercooled scenarios.

We must remark that astrophysical bounds on the  PQ scale $F_a$ require $F_a\gtrsim 10^8$ GeV, that 
is slightly above the scales at which LIGO is most sensitive.
Nevertheless, as we will see,
 the temperature of the phase transition can  be actually slightly smaller than $F_a$ (by up to a factor $\sim 10$), which in the end allows LIGO to probe  these scenarios.
 The capability of LIGO  to probe  PQ phase transitions has  also been pointed out and partially discussed~in~\cite{Dev:2016feu,Dev:2019njv}.

We will also include in our analysis the projected sensitivity for the 
Einstein Telescope (ET) \cite{Punturo:2010zz}. The enhanced sensitivity with respect to LIGO offers the opportunity to probe a much larger area of the parameter space. Therefore ET holds a great promise to probe axion physics.

The article is organized as follows.
In Section~\ref{sensi} we show the sensitivity of Advanced LIGO and the proposed ET
to the parameters $\alpha$, $\beta/H_*$ and $T_*$ of the phase transition.
In Section~\ref{PQmodels} we present the simplest models of  PQ breaking, the KSVZ and DFSZ models, 
analyze their type of phase transitions,
 and  study  their GW signals.
Section~\ref{conclu}  is for conclusions.

\section{Sensitivity of Advanced LIGO and ET  to first-order phase transitions}
\label{sensi}

In this section we show that if  the spontaneous breaking of the PQ symmetry occurred via a first-order cosmological phase transition, then this would have left a stochastic GW signal potentially detectable by LIGO as well as future GW observatories. Indeed, in this case the transition proceeds by bubble nucleation and the collisions between the bubbles as well as the motion of the thermal plasma which surrounds them are sufficiently violent events to generate significant GWs. 
Let us start by reviewing some basic notions that characterize first-order phase transitions and how they can source GW backgrounds.

First-order phase transitions occur when 
there are at least two minima in the scalar potential (which generically depends on the temperature $T$)
and the universe, initially trapped in the minimum with higher energy at high $T$, 
transits to the minimum with lower energy either by thermal fluctuations or quantum tunneling.
In both cases this proceeds at a certain `nucleation' temperature $T=T_n$  through the formation of bubbles of a critical radius $R$ which then expand and percolate.
The rate at which bubbles are produced per unit volume is given by $\Gamma={\cal A} \,e^{-S_B}$
where $S_B$ is the action of the critical bubble, or bounce, and  ${\cal A}$ is a prefactor 
that is usually of order $1/R^4$.
In order for the phase transition to be completed in an expanding universe, we must have $\Gamma\gtrsim H^4$
where $H$ is the Hubble rate.
The nucleation temperature $T_n$ is therefore determined by $\Gamma\sim H^4$ which leads to
\be
S_B(T_n)\sim 4\ln\left(\frac{T_n}{H(T_n)}\right)\equiv S_n\,,
\label{SN}
\ee
where we have taken the approximation ${\cal A}\sim T^4$.
The calculation of $S_B$ depends on the details of the potential and has to be performed case by case.

The  parameters which characterize the first-order phase transition, and which are relevant for the GW signal, 
are the following:
\begin{enumerate}
\item The temperature $T_*$ at the time when the phase transition completes. It can be estimated from energy conservation by equating the latent heat $\Delta V$ (the difference of the potential between the false and true vacuum) plus the energy density in the thermal bath at the nucleation temperature to the energy density of a thermalized plasma, 
$\rho_\gamma(T_*)= \rho_\gamma(T_n) +\Delta V$ with $\rho_\gamma(T)=\pi^2g_*/30\,T^4$. Assuming that the number of relativistic degrees of freedom, $g_*$, does not change much between $T_*$ and $T_n$, one gets
\be
T_*\simeq  
\left(\frac{30}{\pi^2}\frac{\Delta V}{g_*} + T_n^4 \right)^{1/4}\,.
\label{Tstart}
\ee
\item The strength of the first-order phase transition $\alpha$, characterized by the energy density going into the bubbles 
over the thermal energy density of the surrounding plasma:
\be
\alpha=\frac{\Delta V}{\rho_\gamma(T_n)}\,.
\label{alphadef}
\ee

\item The inverse of the duration of the phase transition $\beta=[({d\Gamma}/{dt})/\Gamma]_{T_n}$ \cite{Caprini:2015zlo,Caprini:2018mtu}, which can be approximately determined as
\be
\frac{\beta}{H_*}\simeq T\left.\frac{dS_B}{dT}\right|_{T_n}-4\,,
\label{betadef}
\ee
where 
we have assumed fast reheating so that $H_*\equiv H(T_*)\simeq H(T_n)$, and the $-4$ arises from ${\cal A}\propto T^4$.
\item The bubble wall velocity $v_w$, which is determined by the interaction of the bubble walls with the surrounding plasma. The latter exerts a friction force on the propagation of the walls. In very strong phase transitions ($\alpha\gg1$) one expects that the pressure difference across the bubble walls dominates over the friction of the plasma and bubbles run away, thus $v_{w}\rightarrow 1$,
except in certain cases \cite{Bodeker:2009qy,Bodeker:2017cim}.
In weaker phase transitions ($\alpha\ll1$), we will take the estimate that $v_{w}$ is expected to be close to the speed of sound in the plasma \cite{Steinhardt:1981ct}.

\end{enumerate}

The collisions of bubbles during the phase transition can source GWs of a sizable amplitude. 
Production of GWs in a first-order phase transition has been much discussed previously -- see {\em e.g.} \cite{Maggiore:2018sht,Caprini:2015zlo,Caprini:2018mtu,Caprini:2019egz} for recent reviews. 
The generated GW signal represents a stochastic background and as such it is best characterized by its power spectrum. It is customary to express it in terms of the fraction of the present energy density in GWs per unit decade in frequency, 
\be
\Omega_{\rm GW}(f)=\frac{1}{\rho_c}\frac{d\,\rho_{\rm GW}}{d\,\ln f}\,.
\ee
This signal can be separated into three distinct contributions, 
\be
\Omega_{\rm GW}=\Omega_{\phi}+\Omega_{\rm sw}+\Omega_{t}\,,
\ee 
arising from the collision of the scalar wall profiles, the sound waves in the plasma and from turbulence, respectively. 

The shape and size of each contribution can be estimated separately as reviewed in \cite{Maggiore:2018sht,Caprini:2015zlo,Caprini:2018mtu,Caprini:2019egz}. 
In all cases the power spectrum has a maximum at a characteristic frequency basically determined by the inverse duration $\beta$, and deviates from the maximum by two different power laws. 
In this work we will simply assume the following expressions for the GW spectra as functions of the parameters of the phase transition, quoted in \cite{Caprini:2015zlo,Caprini:2018mtu}:

\begin{itemize}
\item From bubble wall collisions,
\be\label{OmegaPhi}
h^2\Omega_{\phi}(f) = 1.66\cdot 10^{-5}    \left( \frac{H_*}{\beta} \right)^2  \frac{\kappa_\phi^2\, \alpha^2}{(1+\alpha)^2}   
 \left( \frac{100}{g_*(T_*)} \right)^{\frac{1}{3}} \frac{v_w^3}{1+2.4\,v_w^2}
\;\frac{(f/f_\phi)^{2.8}}{1 + 2.8 \, (f/f_\phi)^{3.8}} \,,
\ee
with $h$ the dimensionless Hubble parameter, $\kappa_\phi$ an efficiency parameter which can suppress the contribution from bubble collisions when the effects of the thermal plasma cannot be neglected, and 
the peak frequency today given by 
\be\label{fPeakPhi}	
f_{\phi} = 56.8  \, {\rm Hz} \times 
\left(\frac{\beta/H_*}{10}\right) 
\left(\frac{T_*}{10^8\,{\rm GeV}}\right) 
\left(\frac{1}{1-0.05v_w+0.55\,v_w^2} \right) 
\left( \frac{g_*(T_*)}{100}
\right)^{\frac{1}{6}}~.
\ee

\begin{figure}[t]
\centering
\includegraphics[width=11cm]{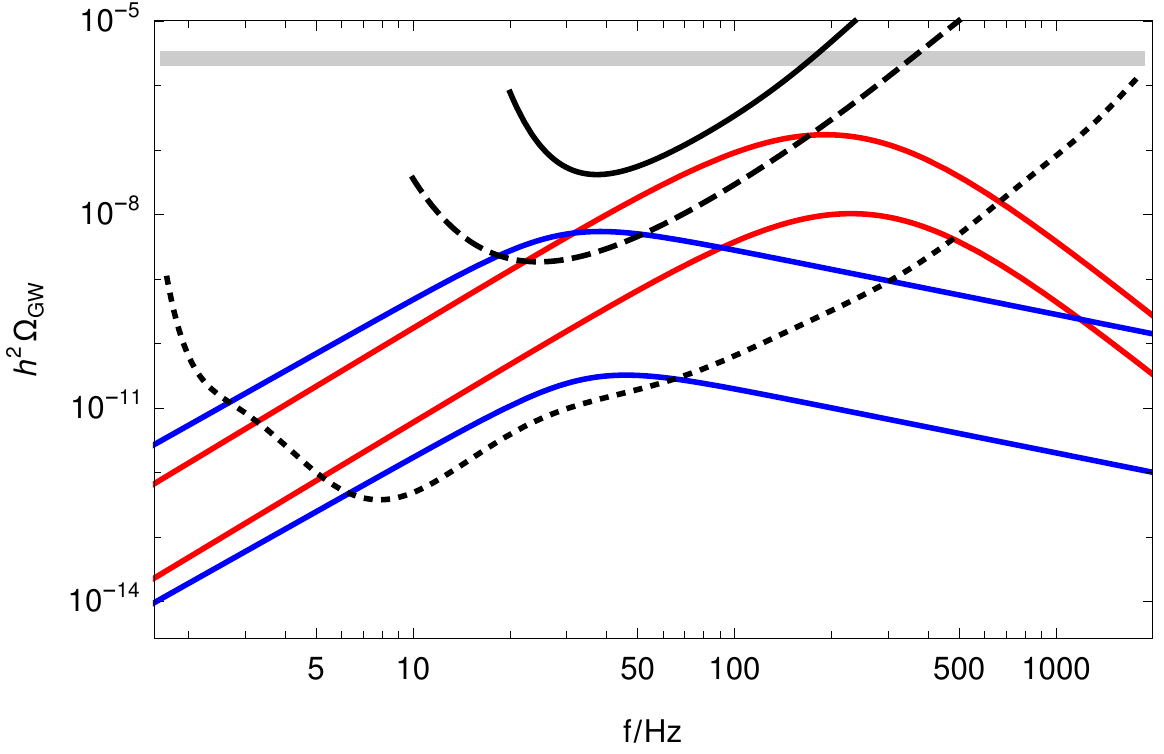}
\caption{\it
Sensitivity curves for stochastic GW searches for LIGO with O1 and O2 data (solid black), for design LIGO (dashed) and for ET (dotted). The thick gray line is the indirect upper bound from Planck CMB data. 
 We also show two representative power spectra 
arising in the PQ model of Sec.~\ref{sub:cw}, corresponding to 
the points `$p_2$' and `$p_3$' of Fig.~\ref{CWcase}.
They have respectively  $F_a=10^8\,$GeV with $\alpha\approx 3.5$, 
and $F_a=10^9\,$GeV with  $\alpha\sim  10^6$. 
We have set $v_w=1$ and show the signals which arise from only bubble wall collisions with $\kappa_\phi=1$ (blue lines) and from only sound waves in the plasma with $\kappa_{\rm sw}=1$ (red lines). 
}
\label{example}
\end{figure}

\item From sound waves in the plasma,
\be\label{OmegaSW}
h^2\Omega_{\rm sw}(f)  =   1.88 \cdot 10^{-5} \, \left( \frac{H_*}{\beta} \right)  \frac{\kappa_{\rm sw}^2 \,\alpha^2}{(1+\alpha)^2} 
 \left( \frac{100}{g_*(T_*)} \right)^{\frac{1}{3}} v_w \;
\frac{ (f/f_{\rm sw} )^{3}}{ \left[1 + 0.75 \,(f/f_{\rm sw})^{2}  \right]^{7/2}}\,, 
\ee
with the peak frequency today given by 
\begin{equation}
f_{\rm sw} = 19  \,  {\rm Hz} \times \frac{1}{v_w} \, \left(\frac{\beta/H_*}{10} \right) 
\left(\frac{T_*}{10^7\,{\rm GeV}}\right) \left( \frac{g_*(T_*)}{100} 
\right)^{\frac{1}{6}}\,.
\label{eq:SWPeak}
\end{equation}

The efficiency parameter $\kappa_{\rm sw}\leq1$ quantifies the fraction of the latent heat which goes into bulk motion. 
Here we shall assume the expression obtained {\em e.g.} in \cite{Kamionkowski:1993fg, Espinosa:2010hh} (see~\cite{Caprini:2015zlo,Caprini:2018mtu} for a recent discussion), which holds for $v_{w}\sim 1$,
\begin{equation}
\label{eq:efficiencysw}
\kappa_{\rm sw}=\frac{\alpha}{0.73+0.083\sqrt{\alpha}+\alpha}\,.
\end{equation}

\begin{figure}[t]
\centering
\includegraphics[width=7.5cm]{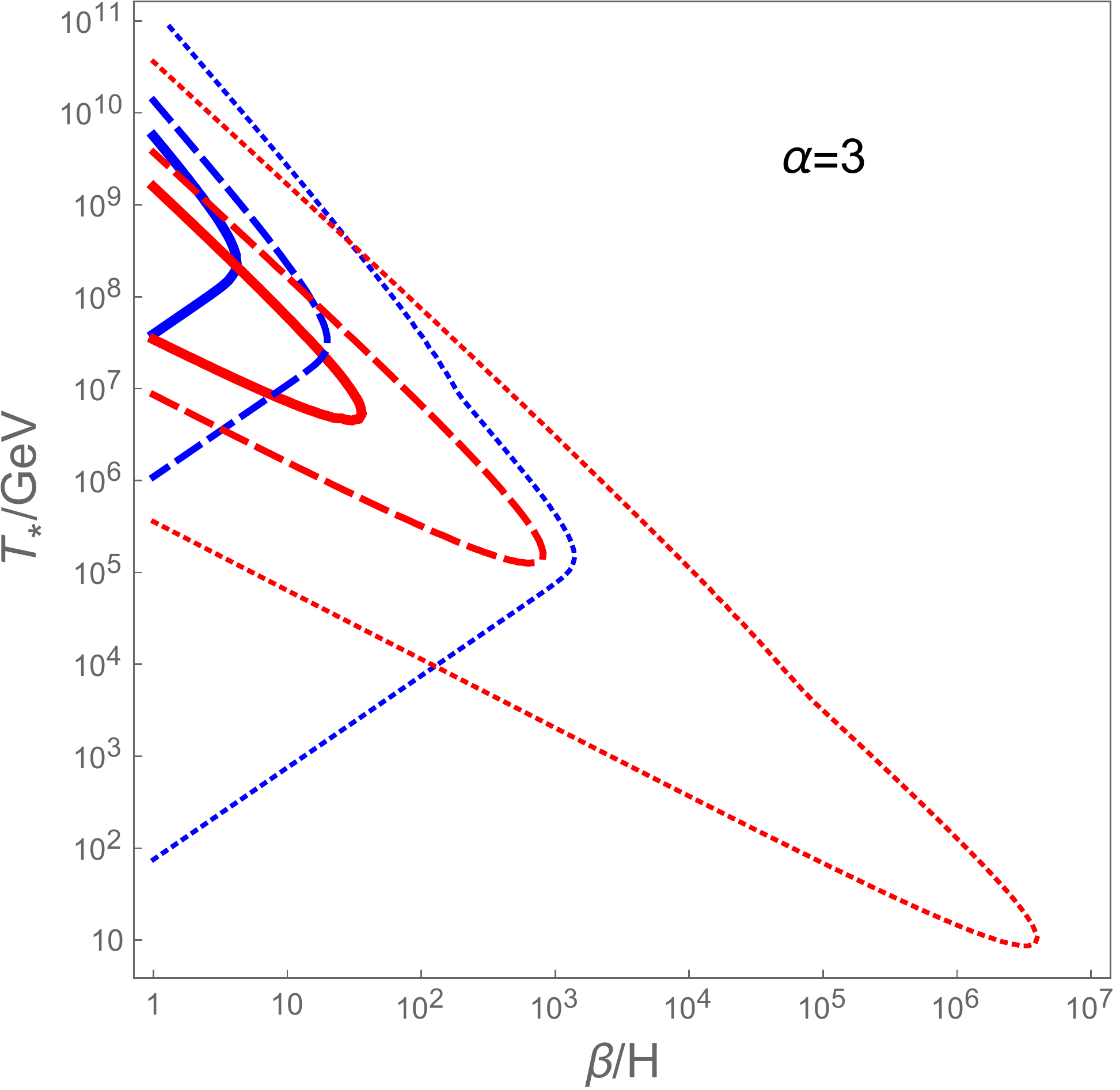}
\qquad
\includegraphics[width=7.5cm]{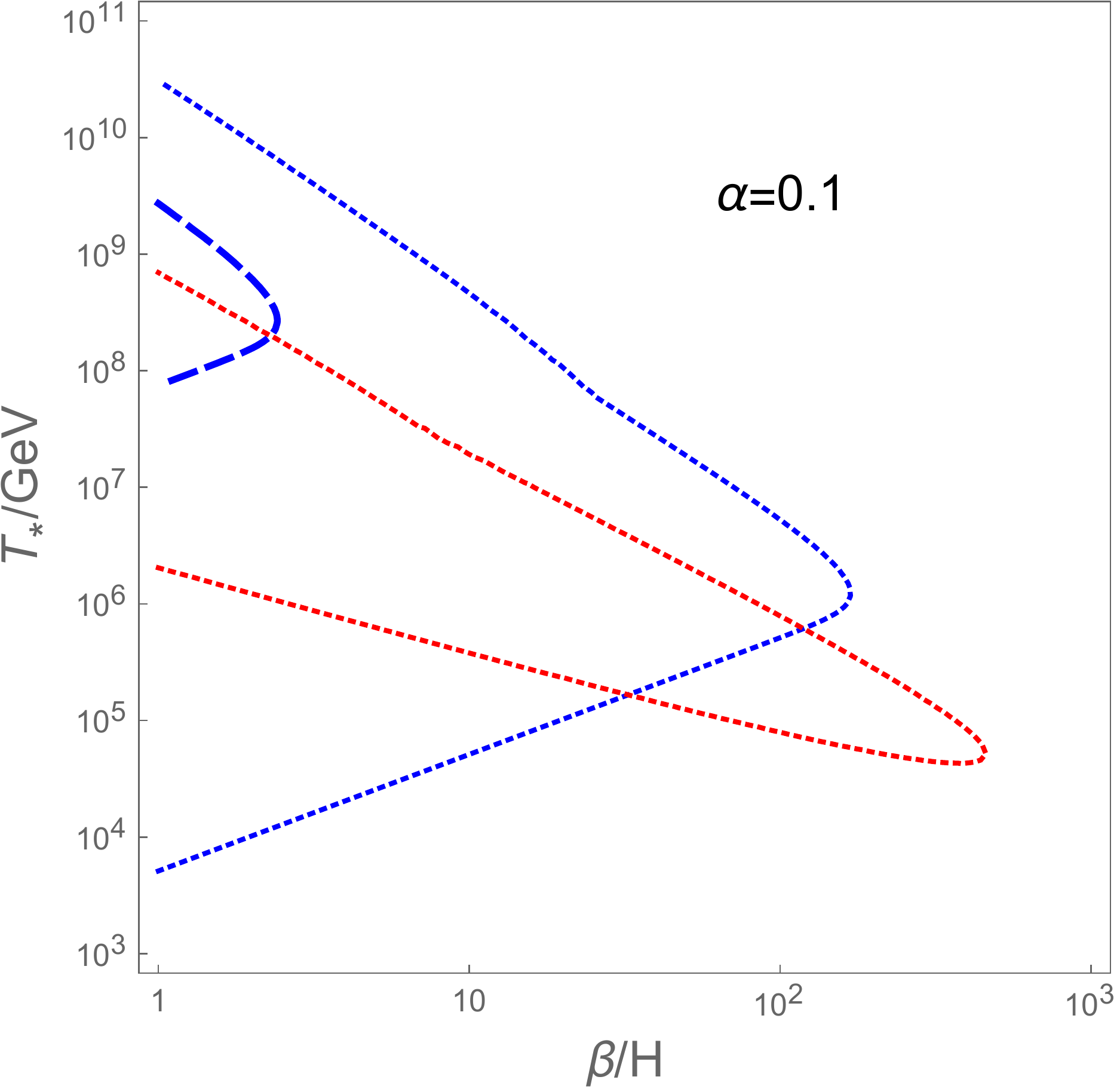}
\caption{\it Sensitivity lines for current LIGO (solid), LIGO at design sensitivity (dashed) and ET (dotted). Blue or red colors refer respectively to whether the signal is mainly sourced by bubble wall collisions (with $\kappa_\phi=1$) or by sound waves   (with $\kappa_{\rm sw}$ given in \eqref{eq:efficiencysw}). 
The points to the left of the curves represent detectable signals. 
Left panel: $\alpha=3$ and $v_w=1$. Right panel: $\alpha= 0.1$ and $v_w=1/\sqrt3$.}
\label{sensitivities}
\end{figure}

\item The contribution from turbulence $\Omega_{t}$ is suppressed (while also being more uncertain), and we will set it to zero for our estimates.
\end{itemize} 
A convenient way to know whether a signal is detectable by a given GW observatory is to compare the power 
spectrum to the so-called {\em power-law integrated curves} \cite{Thrane:2013oya}, which express the sensitivity as the minimal $\Omega$ needed for detection as a function of $f$ (see \cite{Alanne:2019bsm} for an alternative method of presenting sensitivity curves). 
In this work we will be interested in the frequency range which can be probed by ground-based interferometers.
We show in Fig.~\ref{example} the sensitivities of the current Advanced LIGO (with O1 and O2 data \cite{LIGOScientific:2019vic}), as well as the projected sensitivities of the design Advanced LIGO and ET \cite{Punturo:2010zz}. For illustration, we also include in the figure some representative power spectra which arise in the PQ model 
discussed in Section \ref{sub:cw}. 
We also include the indirect limits resulting from CMB data \cite{Pagano:2015hma}. The CMB bound is on the integral $\int df \Omega_{\rm GW}/f$, so how this translates to a bound on the spectral density depends on the shape of the assumed spectrum. This is why we show this bound as a thick line in Fig.~\ref{example}.

\begin{figure}[t]
\centering
\includegraphics[width=7.cm]{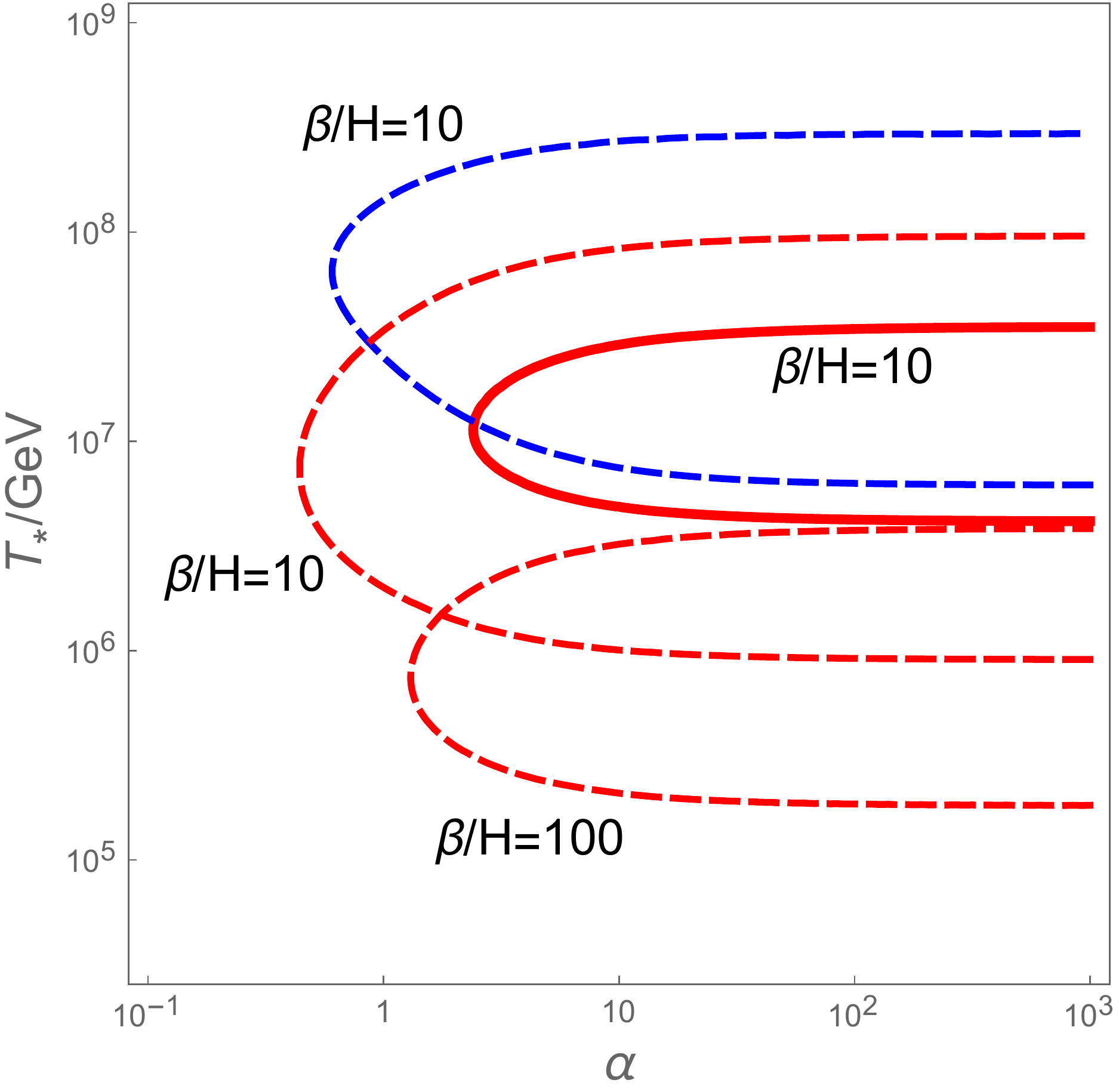}
\caption{\it Sensitivity lines for current LIGO (solid) and LIGO at design sensitivity (dashed) in the $T_* - \alpha$ plane assuming fixed  $\beta/H_*$ equal to 10 and 100 as indicated. Blue or red colors refer respectively to whether the signal is assumed of bubble wall collisions type,  \eqref{OmegaPhi} with $\kappa_\phi=1$ and $v_w=1$, or of sound waves type,  \eqref{OmegaSW} with $\kappa_{\rm sw}$ set by \eqref{eq:efficiencysw}  and  $v_w=1/\sqrt{3}$. 
The points to the right of the curves represent detectable signals.}
\label{Talpha}
\end{figure}

One can then easily obtain which part of the parameter space $(\alpha, \beta, T_*,v_w)$ corresponds to detectable signals by simply checking whether the power spectrum overlaps with the instrument sensitivities. In Fig.~\ref{sensitivities}, we show the resulting detectable regions (to the left of the lines) in the $T_* - \beta/H_*$-plane for the two representative values $\alpha=0.1$ and $\alpha=3$. 
Notice the different shapes of the detectable regions for signals arising from sound waves and bubble wall collisions. For frequencies above the peak in the spectrum, $f \gg f_{{\rm sw}}, f_{\phi}$, the former decays much more rapidly, $\Omega_{\rm sw} \propto f^{-4}$, than the latter, $\Omega_\phi \propto f^{-1}$ (cf.~Fig.~\ref{example}). At the same time, the signal from sound waves increases less rapidly for small $\beta/H_*$, $\Omega_{\rm sw} \propto (\beta/H_*)^{-1} $, than that from bubble wall collisions, $\Omega_\phi \propto (\beta/H_*)^{-2}$.   
Together this causes the lower line limiting the detectable region to have a different slope for the two cases. Furthermore notice that, since the peak frequencies in the spectrum $f_{\rm sw}, f_\phi \propto T_* \beta/H_* $, the tips of the detectable regions move to the lower right in the $T_* - \beta/H_*$-plane if the sensitivity of an instrument increases.
Clearly, for strong first-order phase transitions with $\alpha\gtrsim3$, LIGO at design sensitivity can detect signals that fall into the relevant range for the PQ models,  $T_*\sim 10^7-10^8$ GeV, and it can reach values of $\beta/H_* $ as large as $10^2 - 10^3$. 
Interestingly, even the current O1 and O2 runs of LIGO are capable of ruling out phase transitions  with  $\beta/H_* \lesssim 10$. 
As can be seen in Fig.~\ref{sensitivities},  the improvement on these figures by ET would be rather impressive. 

On the other hand, for small $\alpha$ the possibility to detect a first-order phase transition almost completely fades away at LIGO. This is illustrated in Fig.~\ref{Talpha} where we show the  LIGO sensitivity 
in  the $T_* - \alpha$-plane. 
By taking reasonable values of $\beta/H_* \gtrsim \mathcal{O}(10)$, one clearly sees that in order to possibly detect a signal at LIGO the transition needs to be strong, that is, with $\alpha\gtrsim 1$.
The situation could be  slightly improved with  ET which could reach down to $\alpha\sim 0.1$.

One must be aware that the collection of unresolvable black hole and binary neutron star mergers creates an additional stochastic GW background \cite{LIGOScientific:2019vic}, the so-called `popcorn' background. Given the event rates of these mergers, the magnitude of the popcorn in the LIGO frequency band is around $h^2\Omega\sim 10^{-9}$, which enters in the detectability range for ET and marginally so for LIGO at design sensitivity. This signal represents a 'foreground' for the cosmic GW backgrounds, and it should be subtracted away in order to be able to detect a possible background from cosmological phase transitions. This seems in principle feasible since the power spectra from popcorn and phase transitions differ significantly \cite{LIGOScientific:2019vic}.

It is interesting to note that that  PQ models predict actually two more stochastic sources of GWs in addition to the possible one from the  PQ phase transition. Indeed, since the PQ symmetry is a global $U(1)$ symmetry, it is granted that global cosmic strings will form at the symmetry breaking scale $F_a$. Cosmic string networks radiate GWs, but this is negligible for global strings. Also, at temperatures of order GeV, QCD effects further break $U(1)_{PQ}$ and lead to domain walls, attached to the global strings. The string-wall network then disappears around the QCD scale via rather violent processes where large topological defects collapse and collide. This string-wall network anihilation is similar to a cosmological phase transition and it may give a larger signal.
For QCD axion models the peak frequency of this signal must be around $f\sim 10^{-10}-10^{-7}~\text{Hz}$, which is in the sensitivity range of Pulsar Timing Array observatories.
Unfortunately, the recent numerical simulations of these networks \cite{Saikawa:2017hiv} suggest that the spectrum of this signal falls a bit short to be detectable.

\section{Peccei-Quinn Phase Transition and  its GW signal}
\label{PQmodels}

Having seen the current and future reach of GW interferometers, we now move to the particle physics motivation of this work: the QCD axion solution to the strong CP problem. We start by providing a lightning description of axion physics to set notations, then we investigate the occurrence of a first-order phase transition in the simplest PQ constructions.

Axion models are characterized  by  having a  global  $U(1)_{PQ}$ symmetry
with a $U(1)_{PQ}-SU(3)_c-SU(3)_c$ anomaly.
The $U(1)_{PQ}$ is assumed to be   spontaneously broken by the vacuum expectation value (VEV) of a scalar $\Phi$  at some  scale $F_a$.
The axion $a(x)$  then is the Nambu-Goldstone boson that arises from  this breaking, $\Phi=e^{ia(x)/F_a}F_a/\sqrt{2}+\cdots$.
Due to the $U(1)_{PQ}-SU(3)_c-SU(3)_c$ anomaly, the axion couples to gluons as
\be
\frac{\alpha_s}{8\pi}\frac{a}{F_a}G^{\mu\nu}\widetilde G_{\mu\nu}\,,
\label{definition}
\ee
which leads to a potential for the axion through QCD instantons.
 This  gives $\langle a\rangle=0$, solving the strong CP problem,
and an axion  mass
\be
m^2_a\simeq \frac{m_u m_d}{(m_u+m_d)^2}\frac{m^2_\pi F^2_\pi}{F_a^2}\,.
\ee

We can categorize  PQ models into two different types, depending on the origin of  \eqref{definition}.
Those referred to as KSVZ models \cite{Kim:1979if,Shifman:1979if} contain extra quarks which are responsible for  the anomaly and which generate the term 
\eqref{definition}.
On the other hand, those referred to as DFSZ models \cite{Dine:1981rt,Zhitnitsky:1980tq}  contain extra scalars which,  after being integrated out,
generate the coupling
\be
m_q e^{ia/F_a}\bar q q\,,
\label{massa}
\ee
 where $q$ refers to  SM quarks.  By a  chiral rotation of $q$, the axion can be moved from \eqref{massa} to 
 \eqref{definition}.
Below we discuss the minimal versions of these types of models, their phase transitions and potential GW signals.

\subsection{KSVZ axion models}
The minimal model of this type consists of a scalar $\Phi$  and an extra quark $Q'_{L}$, $Q'_R$ 
with PQ charges $q_\Phi$, $0$ and $-1$,  respectively.
The  interactions, dictated by the PQ symmetry, are given by 
\be
\lambda_\phi( |\Phi|^2-f^2/2)^2+y_{Q'} \Phi^n \bar Q'_L  Q'_R\,,
\ee
where we have set $q_\Phi=1/n$. 
Unfortunately, in this model in which $\Phi$  only interacts with itself and an extra fermion,
the phase transition  is second order, and no significant GWs are expected to be produced from the phase transition.
We could couple $\Phi$ to the SM Higgs, e.g.,  $|H|^2(\kappa |\Phi|^2-\mu^2$).
However,
in order to achieve a viable electroweak (EW) symmetry breaking, we need to tune $\kappa\langle\Phi\rangle^2\approx \mu^2$.
This constraint has not allowed us to find a region of the parameter space where the PQ phase transition is strongly first-order
(see also \cite{Dev:2019njv}). 

We will see later  that supersymmetric versions of the KSVZ model can however have a 
strong first-order phase transition.

\subsection{DFSZ axion models}
\label{DFSZ}

This type of models instead consist of the PQ scalar $\Phi$ and one extra scalar $SU(2)_L$ doublet beyond the one in the SM.
We denote the two doublets as $H_1$ and $H_2$. Their hypercharges are $Y=1$ and $Y=-1$ and we choose their PQ charges as 
$0$ and $-1$, respectively, while the PQ charge of $\Phi$ is $q_\Phi$.
The model should also contain at least one SM quark charged under PQ. 
A minimal option is that only  $u_R$  is charged under PQ, with PQ charge $1$. The  interactions are then fixed by the $U(1)_{PQ}$ symmetry to be 
\be
y_d H_1 \bar Q_L d_R+
y_u   H_2 \bar Q_L u_R+h.c.\,,
\ee
for the quarks in the first family, while the rest of the SM fermions couple only to $H_1$.
The scalar potential is given by
\bea
V&=&\lambda_\phi( |\Phi|^2-f^2/2)^2
+|H_1|^2( \kappa_1 |\Phi|^2-\mu_1^2)
+|H_2|^2( \kappa_2 |\Phi|^2+\mu_2^2)
-(\kappa_3 \Phi^n H_1 H_2+h.c.)\nonumber\\
&+&\lambda_1  |H_1|^4
+\lambda_2  |H_2|^4
+\lambda_{3} |H_1H_2|^2
+\lambda_{4} |H_1|^2|H_2|^2
\,,
\eea
where $n=1/q_\Phi$, $H_1H_2=\epsilon_{ab}H_1^aH_2^b$, and all couplings are real ($\kappa_3$ can be made real by a field redefinition). 
We will for definiteness fix all couplings to be positive in this section. 
For the real parts of the $U(1)_{\rm EM}$-neutral components, $\Phi=\phi/\sqrt{2}$, $H_1=h_1/\sqrt{2}$ and  $H_2=h_2/\sqrt{2}$, we then have
\bea
V&=&\frac{\lambda_\phi}{4}( \phi^2-f^2)^2
+\frac{1}{2}h_1^2( \frac{\kappa_1}{2} \phi^2-\mu_1^2)
+\frac{1}{2} h_2^2( \frac{\kappa_2}{2} \phi^2+\mu_2^2)
-\frac{\kappa_3}{2^{\frac{n}{2}}} \phi^n h_1 h_2\nonumber\\
&+&\frac{\lambda_1}{4}  h_1^4
+\frac{\lambda_2}{4}  h_2^4+
\frac{\lambda_{12}}{4} h_1^2 h_2^2
\label{pot}
\,,
\eea
where $\lambda_{12}=\lambda_3+\lambda_4$.
The mass matrix of $h_{1,2}$ at the PQ-breaking minimum $\phi=f$ is given by
\begin{equation}
{\cal M}^2_H=\left(\begin{array}{cc}\frac{\kappa_1}{2} f^2-\mu_1^2 & -\frac{\kappa_3}{2^{n/2}}  f^n \\ -\frac{\kappa_3}{2^{n/2}}  f^{n} & \frac{\kappa_2}{2} f^2+\mu_2^2\end{array}\right)\,.
\end{equation}
In order to obtain the observed electroweak scale, the determinant of the mass matrix has to be tuned such that\footnote{At the one-loop level, this relation will  of course be modified.}
\begin{equation}
{\rm Det} \ {\cal M}^2_H \sim - m^2_W f^2\ll f^4\,. 
\label{tuningini}
\end{equation}
This is the hierarchy problem which we do not address here but which will be considered below. The SM Higgs is given by the linear combination $H=\cos\theta\, H_1+\sin\theta\, \tilde H_2$ ($\tilde H_2=i\sigma_2H_2^*$) which diagonalizes  ${\cal M}^2_H$ and whose  mass squared is of order $m^2_W$. 
Notice that the mixing angle $\theta$  enters into the expressions for the SM fermion masses: $m_d=y_d \cos\theta\, v/\sqrt{2}$ and $m_u=y_u \sin\theta\, v/\sqrt{2}$.
By integrating out the heavy Higgs doublet, one gets the coupling \eqref{massa}.

The original DFSZ proposal \cite{Dine:1981rt,Zhitnitsky:1980tq} has $n=2$ ($q_\Phi=1/2$) and all three SM up-type quarks are charged under PQ. This choice leads to a cosmological problem~\cite{Zeldovich:1974uw} after the QCD phase transition, since the domain wall number parameter $N_{\text{DW}}$ is larger than one (in particular $N_{\rm DW}=6$ in the original DFSZ proposal). This can be evaded by the introduction of a further small source of explicit breaking of the PQ symmetry~\cite{Sikivie:1982qv}. 
Here instead we make a different choice and focus on $n=1$ ($q_\Phi=1$). In this case, if only the first-family  $u_R$ is charged under the PQ symmetry, we have $N_{\rm DW}=1$ and we avoid the domain wall problem. 
Other choices for the PQ charges and for $n$ will, however, not substantially change our results on phase transitions in these models.

From now on, since $m_{W}\ll f$, we drop the EW scale in our computations. Thus, the tuning \eqref{tuningini} reduces to
\begin{equation}
f^2 \left(\kappa_1 -2\frac{\mu_1^2}{f^2}\right)\left(\kappa_2 +2 \frac{\mu_2^2}{f^2}\right)\simeq 2 \kappa_3^2\,.
\label{tuning}
\end{equation}
In our study of the DFSZ model, we use \eqref{tuning} to fix the parameter $\kappa_{3}$.
The potential \eqref{pot} is then characterized by nine parameters: the scale $f$, the mass parameters $\mu_{1}^2, \mu_{2}^2$, the self-couplings  $\lambda_{1}, \lambda_{2}$ and $\lambda_{\phi}$, the  quartic  couplings $\kappa_{1}, \kappa_{2}, \lambda_{12}$. 
Furthermore, the potential \eqref{pot} is a function of the three scalar fields $h_{1}, h_{2}$ and $\phi$. Nevertheless, 
we will focus on cases where  $h_2$ either vanishes or can be assumed to quickly track its minimum during the phase transition. We will therefore not study its dynamics during the phase transition and only consider its loop effects on the potential for $h_1$ and $\phi$.

It is thus only in the two-dimensional field space of $h_{1}$ and $\phi$ that we will look for a first-order phase transition. In this field space, the potential with signs as chosen in \eqref{pot} can have two minima away from the origin $\textbf{O}$, which we denote with $\textbf{A}$ and $\textbf{B}$, located along the $\phi$ and $h_{1}$ direction respectively (see Fig.~\ref{PTs}):
\begin{equation}
\text{\bf{A}:}\quad  \phi=f,~h_1=h_2=0\ , \quad \quad \text{\bf{B}:}\quad h_1^2=\mu_1^2/\lambda_1, \phi=h_2=0\,.
\end{equation}
Our universe will correspond to the PQ-broken minimum $\textbf{A}$. Therefore, in order to avoid any danger of having an energetically more favorable vacuum at $\bf{B}$, we  require  $V(\textbf{A})<V( \textbf{B})$. This implies the following lower bound on $\lambda_{\phi}$:
\begin{equation}
\label{eq:lower}
\lambda_{\phi}>\frac{1}{\lambda_{1}}\left(\frac{\mu_{1}}{f}\right)^{4}.
\end{equation}
This lower bound is only valid at tree level and can be modified by loop corrections. The point $\bf{B}$ can either be a local minimum or a saddle point of the potential. It is a local minimum (the mass of $\phi$ is positive at $\bf{B}$) if the following upper bound on $\lambda_{\phi}$ is satisfied:
\begin{equation}
\label{eq:upper}
\lambda_{\phi}<\frac{\kappa_{1}}{2\lambda_{1}}\left(\frac{\mu_{1}}{f}\right)^{2}.
\end{equation}
We then find two  possibilities for a strong first-order phase transition in the DFSZ construction
(shown in Fig.~\ref{PTs}):
{\begin{itemize}
\item[${\bf I.}$]
$\bf{O}\rightarrow \bf{A}$, along the $\phi$ direction. The  barrier can be either induced by  thermal corrections,
mainly thanks to the cubic term $T\phi^3$, or by  one-loop corrections of  Coleman-Weinberg type (see \eqref{eq:cw}). 
The latter is more promising for a  strong first-order phase transition, but it requires the mass parameters to be very small 
compared to $F_a$.
\item[${\bf II.}$] $\bf{O}\rightarrow \bf{B}\rightarrow \bf{A}$, first along the $h_{1}$ direction and later along some $\phi-h_1$ trajectory.
If \eqref{eq:upper} is satisfied, a tree-level zero-temperature barrier separates the minima $\bf A$ and $\bf B$ which can lead to a first-order phase transition in the second step.  In this case the universe goes through an intermediate  
phase with a large EW symmetry~breaking~scale.

\end{itemize}
We explore the two possibilities above in the following subsections Sec.~\ref{sub:cw} and Sec.~\ref{sub:twostep}, respectively.

\begin{figure}[t]
\centering
\includegraphics[width=0.5\textwidth]{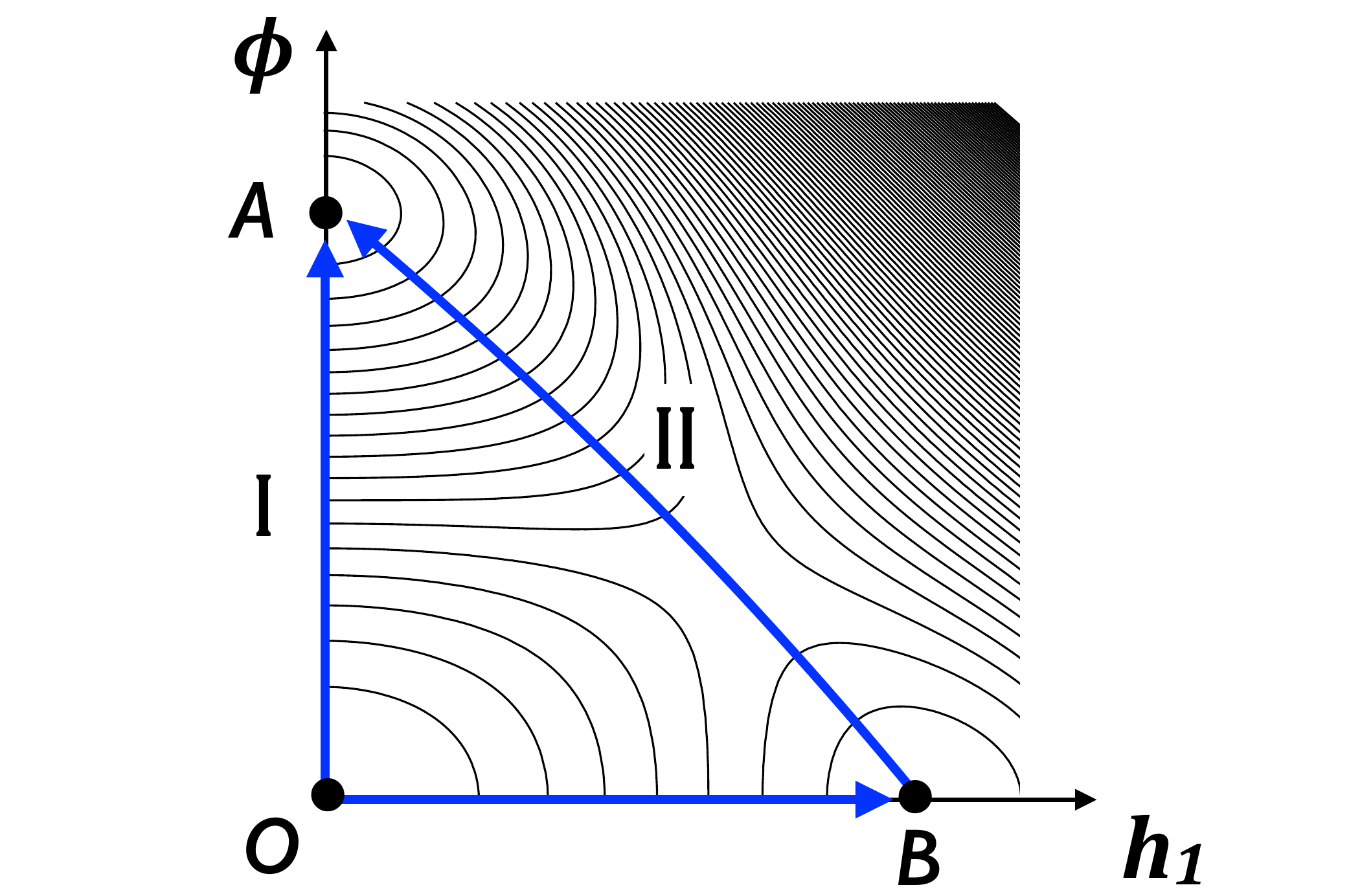}
\caption{\it  Field trajectory $\bf I$ and $\bf II$ of the phase transition.}
\label{PTs}
\end{figure}

\subsubsection{Thermal and Coleman-Weinberg driven first-order phase transition}
\label{sub:cw}

Let us  consider the phase transition in the direction of $\phi$ (trajectory {\bf I} in Fig.~\ref{PTs}).
To ensure that $h_1$ stays zero
 during the phase transition, we roughly need $T_n \gtrsim \mu_1$ for signs as chosen in \eqref{pot} (otherwise one first rolls/tunnels towards the $h_1$-direction, leading to a trajectory like {\bf II} in Fig.~\ref{PTs}). 
This limits the smallest $T_n$ that is achievable. The smaller $T_n$, however, the stronger is the GW signal as we will see below. Another option is  to flip the signs of both $\kappa_1$ and $\mu_1^2$ in \eqref{pot}. One can show that if $\mu_1^2$ is chosen sufficiently large, a tachyonic direction in $h_1$ and $h_2$ only develops for $\phi$ very close to its minimum. Both fields can therefore be consistently set to zero and their dynamics ignored during the phase transition.\footnote{One may worry that negative $\kappa_1$ can lead to a runway direction in the potential. In order to avoid this, one needs to impose that $|\kappa_1| < 2 \sqrt{\lambda_\phi \lambda_1}$. A natural value for $|\kappa_1|$ can be estimated from the two-loop contribution involving 
 $\kappa_2$  and the gauge couplings. Using this estimate, we find that this condition can be expected to be fulfilled.} We will further assume that all couplings to $h_1$ are sufficiently small and it is sufficiently heavy that we can also ignore its loop-corrections to $\phi$ and $h_2$. We are then left with $\phi$ and $h_2$, where the latter affects the dynamics of the phase transition only via loop corrections. The resulting potential for $\phi$ at loop-level and for finite temperatures is discussed in Appendix~\ref{app1}.

A first-order phase transition can occur due to a thermal barrier generated by the cubic term $\sim T\phi^3$,
mostly arising from loops of $h_2$. Nevertheless, when the daisy masses are included (see  Appendix~\ref{app1}), this cubic term
is diminished and the barrier is usually small~(see e.g.~\cite{Quiros:1994dr} and \cite{Curtin:2016urg} for a recent discussion).
The resulting values of $\alpha$ are then small and those of $\beta/H_*$  large, leading  only to a weak GW signal. 
This can be seen for example for  the point marked by $p_1$ in Fig.~\ref{CWcase}, 
  calculated with $\kappa_2=2$, $\lambda_\phi \sim 10^{-2}$ and $\sqrt{\lambda_\phi} f \sim 10^6\,$GeV. As can be seen in the plot, the phase transition for this case has $\beta/H_* \sim 100$, while $\alpha \sim 0.2$. Note that the barrier in this case already has a contribution from the Coleman-Weinberg corrections which we discuss below. A purely thermal barrier would have even larger $\beta/H_*$ and smaller $\alpha$.

A second more promising possibility  for a strong first-order phase transition arises in the limit 
in which the mass parameters are small, $\mu^2_2, \lambda_\phi f^2\ll  f^2$.
In this case, the  $T=0$  potential for $\phi$ 
 becomes almost scale invariant and can be written as 
\be
V=\frac{1}{4}\lambda_\phi(\phi) \phi^4\,.
\label{CWP}
\ee
Due to one-loop corrections, $\lambda_\phi(\phi)$ depends logarithmically on $\phi$ and the potential is thus of Coleman-Weinberg type \cite{Coleman:1973jx}
(when the logs are large, this potential must be RG-improved). If $\lambda_\phi(\phi)$ is negative for small $\phi$ and turns positive for large $\phi$, a minimum develops close to where the coupling crosses zero. More precisely, the minimum is determined by
\be
\lambda_\phi (\langle\phi\rangle )=-\frac{1}{4}\beta_{\lambda_\phi} (\langle\phi\rangle)\,,
\label{minCW}
\ee
where $\beta_{\lambda_\phi}=d\lambda_\phi/d\ln\phi$. Notice that now $F_a\equiv \langle\phi\rangle\not= f$.
Considering only the couplings $\lambda_\phi$ and $\kappa_2$,   we have 
 \be
 \beta_{\lambda_\phi}=\frac{\kappa_2^2}{8\pi^2}+\frac{5\lambda_\phi^2}{4\pi^2}\ ,\ \ \ \ \ \ 
\beta_{\kappa_2}=\frac{\kappa_2^2}{4\pi^2}+\frac{\kappa_2\lambda_\phi}{2\pi^2}\,.
\label{betas}
\ee
From \eqref{minCW}, we can fix one parameter, say $\lambda_\phi$, 
and therefore we are left with only  one  free coupling, $\kappa_2$.
Using \eqref{minCW}, we obtain at the minimum
\be
V_{\rm min}=-\frac{1}{16}\beta_{\lambda_\phi}(\langle\phi\rangle)\langle\phi\rangle^4\simeq -\frac{\kappa_2^2}{128\pi^2}F^4_a\,.
\label{potCWmin}
\ee
The phase transition of Coleman-Weinberg models with a potential given by \eqref{CWP}  
was first studied in  \cite{Witten:1980ez}.\footnote{See \cite{Brdar:2018num} for an earlier study of GWs in the LIGO frequency band which originate from the phase transition of a Coleman-Weinberg model.}
Let us  sketch here how this proceeds.
When  non-zero temperature effects are included, the potential at small $\phi$  is always dominated by  thermal corrections which lead to
\begin{equation}
\label{eq:mphicw}
V_T=D_\phi T^2\phi^2+\cdots\,,
\end{equation}
where $D_\phi$ is given in \eqref{dphi}.
Therefore at any non-vanishing temperature, the curvature of the potential is always positive near $\phi=0$ and this point is a (local) minimum.  In fact, at very high temperatures, the thermal corrections are so large that the  minimum \eqref{minCW} at $\phi=F_a$   is lifted, and the point $\phi=0$ is the only minimum of the potential.
This implies that at a certain temperature  $T_{c}$,  the two minima are degenerate,  and 
it becomes favorable to tunnel from $\phi=0$ to  $\phi=F_a$. Notice that the barrier is generated thanks to $\lambda_\phi$
being negative for $\phi\leq \langle\phi\rangle$. 

As was discussed in \cite{Witten:1980ez}, $O(3)$-symmetric bubbles dominate tunneling in this case and in the limit of small temperatures $T$ their action is well approximated by
\be
S_B =\frac{S_3}{T}\simeq 18.9\frac{\sqrt{2 D_\phi}}{-\lambda_\phi(T)} \simeq 7.7\frac{\sqrt{\kappa_2(T)+2 \lambda_\phi(T)}}{-\lambda_\phi(T)}\,.
\label{bounceCW}
\ee
From this, we see that $S_B$ can slowly evolve from large values to small values, since
$-\lambda_\phi(T)$  grows as  $T$ decreases.
This can eventually allow the criterion in \eqref{SN} to be satisfied and thus the phase transition to happen 
at some temperature $T_n\ll F_a$.
While  trapped  in the false vacuum, the universe inflates  with $H^2=\Delta V/(3M_P^2)$  and supercools.
We can calculate $T_n$ using  \eqref{SN} where now
\be
S_n\simeq 4\ln\left(\frac{T_n M_P}{\sqrt{\Delta V}}\right)\simeq 4\ln\left(\frac{8\sqrt{2}\pi}{\kappa_2}\frac{T_n M_P}{F_a^2}\right)\,.
\ee
From \eqref{betas}, we see that the smaller $\kappa_2$, the slower does $-\lambda_\phi(T)$ grow with decreasing $T$ and therefore the more supercooling we have.
Notice that there is a lower bound for $T_n$, since $S_n$ also decreases with $T_n$ and at some point becomes too small and $S_B$ can never reach its value.

Due to this (long) period of   inflation, where the temperature drops exponentially, the thermal plasma  is diluted,  and we have 
$\alpha\gg 1$. 
Furthermore, from \eqref{betadef} and \eqref{bounceCW}, we obtain
\be
\frac{\beta}{H_*}\simeq \frac{\beta_{\lambda_\phi}(T_n)}{-\lambda_\phi(T_n)} S_n-4\,.
\label{betaCW}
\ee
We can now see under which conditions a slow transition can be achieved.
In principle,  since $\beta_{\lambda_\phi}$ in \eqref {betaCW} 
is one-loop suppressed,  one would expect 
that ${\beta}/{H_*}\sim 1$ can be easily achieved.
However, also $-\lambda_\phi(T_n)$ is one-loop suppressed near the minimum as follows from \eqref{minCW}. In order to make it larger than that, one needs $T_n\ll F_a$.
To be more explicit, let us consider the one-loop coupling
$\lambda_\phi(\phi)\sim  -\beta_{\lambda_\phi}\ln F_a/\phi$.
We then roughly obtain from \eqref {betaCW}
\be
\frac{\beta}{H_*}\sim
\frac{4}{\ln F_a/T_n}\ln\left(\frac{T_n M_P}{F_a^2}\right)-4\,,
\ee
which reaches values of order one at $T_n\ll F_a$. 

Having $\alpha\gg 1$ and the possibility of $\beta=O(1)$,
this scenario then can lead to a maximal signal in GWs,
which we expect to be mainly sourced by the collision of runaway bubble walls themselves since supercooling exponentially dilutes the thermal plasma around them.
 From \eqref{Tstart} together with  \eqref{potCWmin}, we can relate  $T_*$ to $F_a$:
\be
T_*\simeq 0.1\sqrt{\kappa_2}\, F_a\,.
\label{TstartCW}
\ee
This predicts $T_*$ to be slightly below $F_a$, making LIGO and ET (see Fig.~\ref{sensitivities}) 
quite  suited to test the interesting region $F_a\sim 10^{8}-10^{10}$ GeV. 

\begin{figure}[t]
\centering
\includegraphics[width=0.6\textwidth]{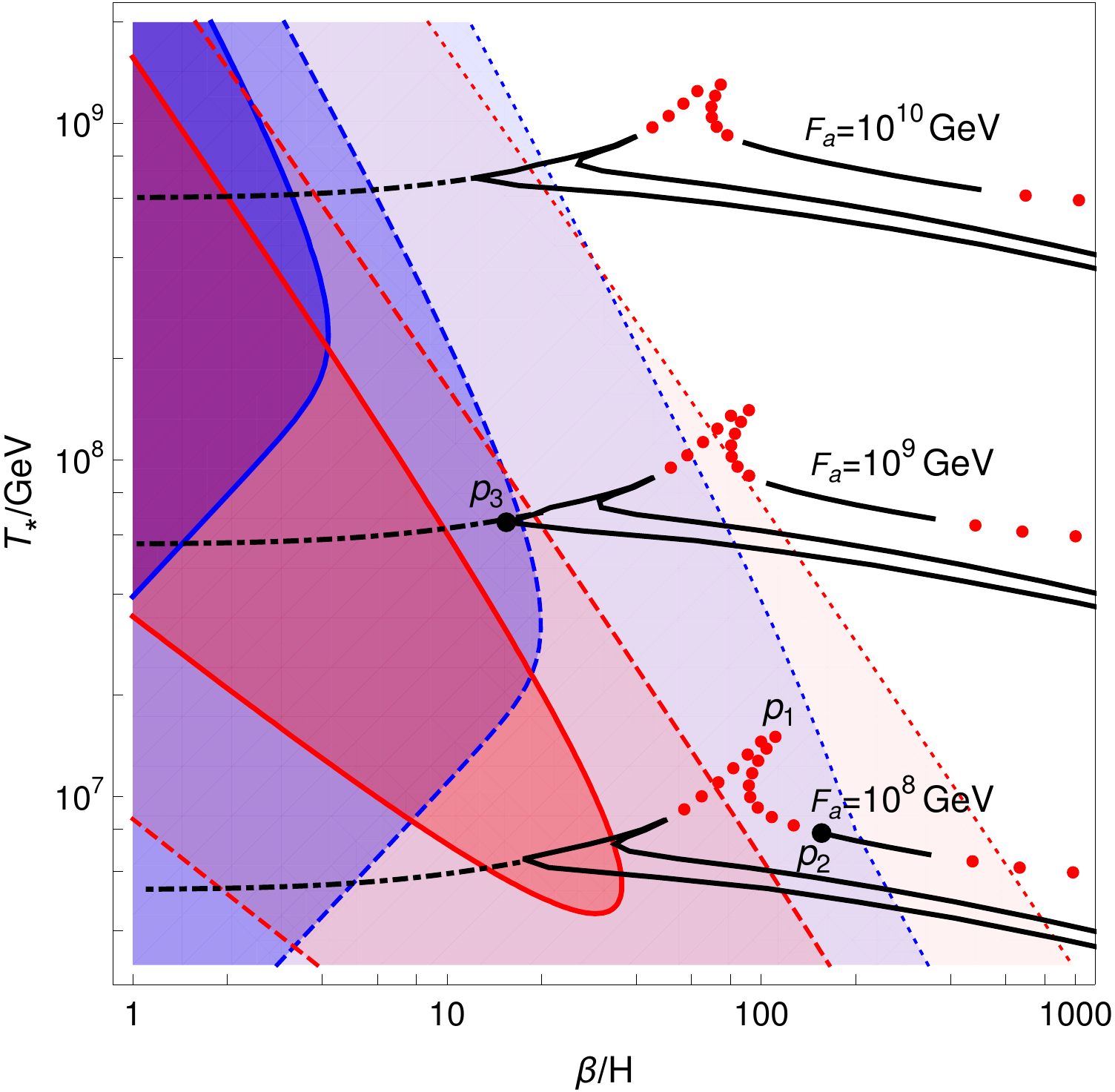}
\caption{\it  Predicted values of $T_*$ and $\beta/H_*$ for 
the DSFZ axion model \eqref{pot} for $F_a=10^8$, $10^9$ and $10^{10}$  {\rm GeV} (see Section~\ref{sub:cw} for more details).
The  present and expected future experimental GW reaches are depicted as red and blue areas, using the same color code as in Fig.~\ref{sensitivities}.}
\label{CWcase}
\end{figure}

We have calculated the properties of the phase transition by numerically solving the bounce equation for the potential \eqref{pot} plus its thermal and Coleman-Weinberg corrections as discussed in Appendix~\ref{app1}. We have performed this calculation for both $O(3)$- and $O(4)$-symmetric bubbles and have confirmed that the former indeed dominate. The resulting values of $T_*$ and $\beta/H_*$
for $F_a=10^8$, $10^9$ and $10^{10}$  {\rm GeV} are shown in Fig.~\ref{CWcase}. We have chosen the two relevant mass parameters, $\mu_2$ and $\sqrt{\lambda_\phi} f$ (where $\lambda_\phi$ is the tree-level coupling), equal for concreteness and hierarchically smaller than $F_a$ in order to be in the Coleman-Weinberg regime. 
Furthermore, we have fixed $\lambda_\phi$ as discussed above and scanned over different values of $\kappa_2 \leq 2$. This gives rise to the solid lines in Fig.~\ref{CWcase} which for each $F_a$ from top to bottom correspond to $\mu_2=\sqrt{\lambda_\phi} f= (10^{-2}$,$10^{-3}$,$10^{-4}$) $F_a$, respectively. By decreasing $\kappa_2$, one moves along these lines towards smaller $T_{*}$ (as is expected from \eqref{TstartCW}). As follows from the discussion above, as long as $T_n \gg \mu_2 ,\sqrt{\lambda_\phi} f$, we have that $T_n$ and $\beta/H_*$  decrease if one lowers $\kappa_2$. This regime corresponds to the parts of the lines in Fig.~\ref{CWcase} with positive slope. Eventually, however, one reaches $T_n \sim \mu_2 ,\sqrt{\lambda_\phi} f$. Since we have chosen the mass of $\phi$ to be tachyonic (cf.~\eqref{pot}),
this mass compensates the thermal barrier (cf.~\eqref{eq:mphicw}) at lower temperatures and the phase transition thus always happens at $T_n \sim \sqrt{\lambda_\phi} f$ if one lowers $\kappa_2$ further. Since this removal of the barrier happens rapidly at around $T_n \sim \sqrt{\lambda_\phi} f$, $\beta/H_*$ then begins to grow again for decreasing $\kappa_2$. This regime corresponds to the parts of the lines in Fig.~\ref{CWcase} with negative slope. We thus find that for every given hierarchy between $F_a$ and $\mu_2,\sqrt{\lambda_\phi} f$, there is a minimal $\beta/H_*$ that can be reached. 
Furthermore, the  dash-dotted lines in Fig.~\ref{CWcase} show results of an analytical approximation following \eqref{bounceCW},  \eqref{betaCW} and \eqref{TstartCW} for the case $\mu_2=\sqrt{\lambda_\phi} f=0$. As expected, this case allows to reach much lower values of $\beta/H_*$. Note also that the solid lines only delimit points with $\alpha \geq  3$, while some representative points with $\alpha < 3$ are shown in red. The values of $\alpha$ always increase on the parts of the lines with positive slope, while they eventually decrease again on the parts with negative slope. 
The restriction to $\alpha \geq 3$ was made since the amplitudes of the GWs becomes independent of this parameter in the limit of large $\alpha$ (see \eqref{OmegaPhi} and \eqref{OmegaSW}). 

In Fig.~\ref{CWcase}, the current and expected reaches of the GW observatories are then shown for $\alpha=3$. Since the amplitudes of the GWs increase by about $40\%$ when going from $\alpha=3$ to very large $\alpha$, the true reaches in the very supercooled regime are slightly higher than what is shown.  
Solid lines delimit sensitivity regions for current LIGO, dashed ones for LIGO at design sensitivity and dotted ones for ET. We expect that in the very supercooled regime of the DFSZ axion model, GWs are dominantly produced by bubble collisions. The sensitivity regions for this case are shown in blue (setting $v_w=1$ and $\kappa_\phi=1$).  
For less supercooling (as expected in particular for the points with small $\alpha$), sound waves can instead be the main source of GWs. We plot the sensitivity regions for this case in red (setting $v_w=1$ and $\kappa_{\rm sw}=1$).
We see from Fig.~\ref{CWcase} that  part of the parameter space could be already detected at LIGO, while other parts will have to wait for ET. The power spectra for the points marked as `$p_2$' and `$p_3$' in Fig.~\ref{CWcase} are shown in Fig.~\ref{example}.

\subsubsection{Cooled two-step phase transition}
\label{sub:twostep}

We now focus on the case of a two-step first-order PQ phase transition,  along the trajectory $\bf{II}$ in Fig.~\ref{PTs}.
Let us first understand under what circumstances the two-step phase transition can occur and be strong enough to source a detectable GW signal. 
From Fig.~\ref{Talpha}, it is clear that LIGO can probe only transitions with $\alpha>1$. In the case of a standard two-step transition, where the minimum $\bf{B}$ develops at a temperature $T_{h_1}$ which is higher than the temperature $T_{\phi}$ at which the PQ minimum appears, these values of $\alpha$ are difficult to obtain. Indeed in this situation the universe cannot cool much if the transition is to be completed, since the barrier between the two minima is already present at tree level. This is in contrast with the previously discussed Coleman-Weinberg driven scenario.

However, in the DFSZ scenario a new possibility arises: namely, that $T_{\phi}> T_{h_1}$, but that below $T_{\phi}$ the universe is stuck for a while at the origin, due to a loop-induced barrier which opposes rolling/tunneling along the $\phi$ direction. In this case, a two-step transition can occur, as below $T_{h_1}$ the universe tracks the local minimum in the $h_{1}$ direction (second order/crossover phase transition). If $T_{h_1}$ is sufficiently small, large values of $\alpha$ are obtained whenever the transition can complete. For this reason, here we focus on this cosmological history. 

We already know of one way to realize this: that is, to make use of the Coleman-Weinberg induced barrier in the $\phi$ direction. Alternatively, a barrier induced by  $\phi^3 T$ terms    arising from thermal loops may also suppress tunneling, although it requires large values of $\kappa_{2}$. In both cases, the crucial ingredient which is peculiar to the DFSZ scenario is the presence of extra bosonic fields coupled to $\phi$, beyond the content of the doublet $H_{1}$. For concreteness, here we focus on the case in which tunneling along $\phi$ is suppressed because of the barrier induced by Coleman-Weinberg corrections due to $h_{2}$ loops. We then discuss the values of $\lambda_{\phi}$ and $\kappa_{1}$ which allow for this scenario to occur, while we keep the rest of the parameters fixed as follows. Since $T_{h_1}\sim \mu_{1}/\sqrt{D_{h_{1}}}$ (where $D_{h_1}$ is defined in \eqref{dh1}), we take $\mu_{1}\lesssim 0.1 f$ to ensure that $T_{h_1}\ll f$. 
Also  we take $\mu_{2}=0.1 f$ and $\kappa_{2}\sim 1$.
 Furthermore, $\lambda_{1}$ is related to the SM Higgs quartic coupling,\footnote{Below the heavy Higgs doublet mass, the SM quartic is given by 
$\lambda_{\rm SM}=\lambda_1\cos^4\theta+\lambda_2\sin^4\theta+\lambda_{12}\cos^2\theta\sin^2\theta$. In addition,
integrating out the heavy singlet $\phi$ gives an extra contribution $\Delta\lambda_{\rm SM}=-\kappa_{\rm SM}^2/(2 M_\phi^2)$
where $\kappa_{\rm SM}$ and $M_\phi$ are respectively  the coupling of $\phi$ to the SM Higgs and its mass.}
which at the energies  we consider is of order $0.01$.
For this reason we take $\lambda_{1}\gtrsim 0.01$.

A local minimum in the $h_{1}$ direction occurs if the upper bound \eqref{eq:upper} on $\lambda_{\phi}$ is respected. For $\lambda_{\phi}\lesssim 10^{-3}$, this is easily satisfied and the potential in the $\phi$ direction is dominated by Coleman-Weinberg corrections due to $h_2$. This also ensures that the tree-level lower bound on $\lambda_{\phi}$ is relaxed, as the minimum $\bf{A}$ is always the global minimum of the potential. Interestingly, completion of the transition from $\bf{B}$ to $\bf{A}$ is facilitated in this case, since the minima are always significantly non-degenerate.

\begin{figure}[t]
\centering
\includegraphics[width=0.55\textwidth]{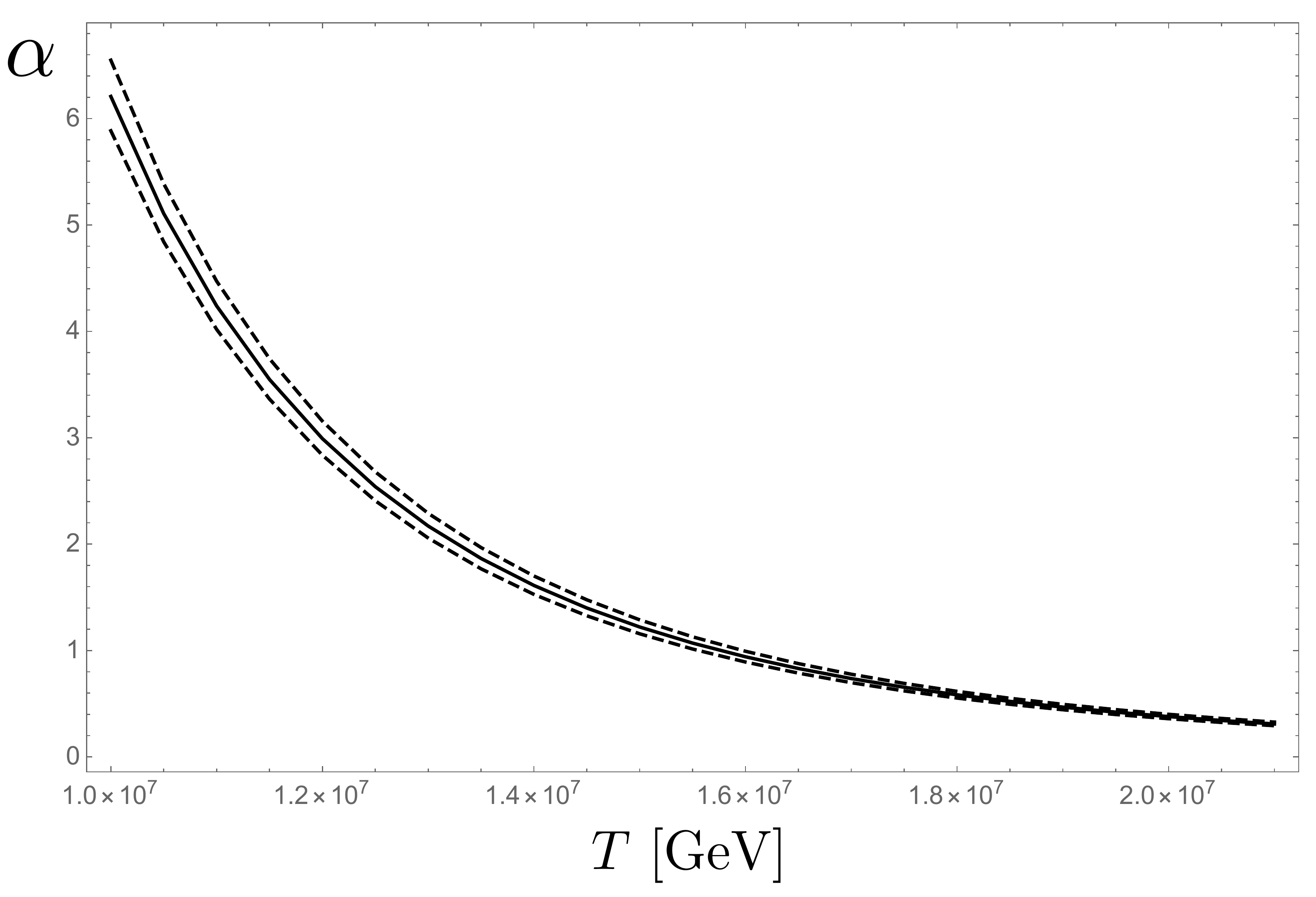}
\caption{\it Latent heat parameter $\alpha$ as a function of temperature for representative choice of parameters with $T_{h_1}\simeq 2.2\cdot 10^{7}~\text{GeV}$. In order to produce this plot, we have fixed $\mu_{1}=0.09f, \lambda_{1}=0.05, \kappa_{2}=1.5, \mu_{2}=0.1f$ and $f=10^8$ GeV. We also fixed $\kappa_{1}=3\mu_{1}^{2}/f^{2}$, since the dependence of $\alpha$ on this parameter is very mild. The solid, dashed upper and dashed lower lines are obtained respectively for $\lambda_{\phi}=10^{-3},~10^{-4},~2\cdot10^{-4}$. Very similar curves can be obtained for smaller values of $\mu_{1}$ and $\lambda_{\phi}$, starting at smaller values of $T_{h_1}$, thus larger values of $\alpha$.}
\label{fig:Latent}
\end{figure}

We then proceed to a numerical investigation of the parameter space for this type of two-step transition.  As mentioned above, even though the potential is a function of three fields, we can focus on the dynamics of $\phi$ and $h_{1}$ only. 
The rest of the fields 
of the DFSZ model will only affect the potential of $\phi$ and $h_{1}$ at the loop level.
These are  all components of the doublets $H_{2}$ and $H_{1}$, the imaginary part of $\Phi$, 
the EW gauge bosons and the top quark.
 We fix $\mu_{1}=0.09f, \mu_{2}=0.1f, \lambda_{1}=0.05, \lambda_{2}=0.01, \lambda_{12}=10^{-3}$ and the gauge couplings as well as the top Yukawa coupling to $0.6$, as appropriate for $f\sim 10^{8}-10^{10}~\text{GeV}$. Finally, in order to consider interesting frequencies of the GW signal, we fix $f=10^{8}~\text{GeV}$. For $f\gtrsim 10^{9}~\text{GeV}$, the transition necessarily requires very small values of $\beta/H_{*}$ to be detectable by LIGO and/or~ET.

We vary $\lambda_{\phi}$ and $\kappa_{1}$ while requiring that tunneling along the $\phi$ direction does not occur until at least $T_{h_1}$. We find that this condition is respected for any value of $\kappa_{1}$, as long as $\lambda_\phi\lesssim 0.002$. For values of $\kappa_{1}$ close to the lower bound $\kappa_{c}=2 \mu_{1}^{2}/f^{2}\simeq 0.02$, the local minimum $\bf{B}$ appears at $T_{h_1}\simeq 2\cdot 10^{7}~\text{GeV}$, while $T_{\phi}\sim 5\cdot 10^{7}~\text{GeV}$. 

We show the evolution of the latent heat parameter $\alpha$ for temperatures below $T_{h_1}$ in Fig.~\ref{fig:Latent}  for representative choices of parameters $\lambda_{\phi}$ and $\kappa_{1}$. It is clear that $\alpha\gtrsim 1$ can be obtained with these choices of parameters if there is  just a  mild cooling of $\sim 20~\%$, {\it i.e.}, if $T_n\lesssim 0.8\, T_{h_1}$.
 Alternatively, one can consider smaller values of $\mu_{1}, \mu_{2}$ and $\lambda_{\phi}$, according to \eqref{eq:upper}. In this way $T_{h_1}$ can be made smaller, therefore ensuring that values of $\alpha$ above one are obtained even when the universe  immediately tunnels below $T_{h_1}$.

Tunneling from $\bf{B}$ to $\bf{A}$ is numerically investigated by means of the multi-field tunneling package \texttt{AnyBubble}~\cite{Masoumi:2016wot}. We find, as expected, that $O(3)$ bubbles only provide a closed window for tunneling to occur: namely, the tunneling action $S_{3}/T$ initially decreases as the difference in vacuum energy of the two minima slightly increases (because the PQ minimum becomes deeper), then reaches a minimum value after which it grows again rapidly (because $\Delta V_T$ remains constant (and then $S_3\approx \text{constant}$), while the temperature keeps decreasing
(and then $S_{3}/T$ becomes larger)). For values of $\lambda_{\phi}$ and $\kappa_{1}$ close to the line determined by the upper bound \eqref{eq:upper}, we find that tunneling occurs very rapidly below $T_{h_1}$, with $\alpha\simeq 0.2$ and $\beta/H_{*}\gg 10^{2}$, as expected since in this region the tree-level barrier is small. However, as we move away from this limit, we find points in parameter space where $T_{n}\simeq 1.5\cdot 10^{7}$ and $\alpha\gtrsim 1$. For these points, we also find $\beta/H_{*}\lesssim 100$, since the transition occurs only after some cooling.
These values are enough to make the associated GW signal detectable at ET independently of the main source of GWs and even at design LIGO, if sound waves are the dominant source of GWs. While we leave a detailed numerical scan of the values of $\beta/H_*$ in the parameter space of the model for future work, we expect that small regions with $\beta/H_*\lesssim 10$ should arise as we move further away from the upper bound \eqref{eq:upper}, close to the region in which the universe remains stuck in $\bf{B}$ forever.\footnote{Here we have not considered tunneling due to $O(4)$ bubbles. We also expect that there is  a small region of parameter space where $O(4)$ tunneling can occur at low temperatures, with larger values of $\alpha$, when $O(3)$ tunneling is inefficient.} This would open up the possibility to detect the signal at LIGO, independently of the specific source of GWs.

In this latter respect, our two-step  PQ phase transition may be characterized by a further peculiarity. Indeed, for $\alpha\gtrsim 1$ it is not clear whether bubbles can achieve a runaway regime, nor whether the main source of GWs is the collisions of the walls or the sound waves in the thermal plasma, or in fact an admixture of both. Since our transition involves the EW gauge bosons, one should consider the implications of transition radiation~\cite{Bodeker:2017cim} as these particles change mass  across the bubble walls. However, in our case the EW symmetry is initially broken at \textbf{B}, with gauge bosons receiving masses $m_{W}\sim \mu_1$ in the second-order transition from the origin to \textbf{B}. In the first-order transition from $\bf{B}$ to $\bf{A}$ the gauge bosons become light, which is the opposite of the case discussed in~\cite{Bodeker:2017cim}.
Therefore, in our case it should be possible for bubbles to run away even if they are surrounded by a thermal plasma, which would lead to  $v_{w}\simeq 1$ and a GW signal sourced by both sound waves and bubble collisions. Having an early phase of broken EW symmetry, with very massive gauge bosons at high energies, may also lead to interesting possibilities for baryogenesis at high scales. We leave the interesting questions above for future work.

\subsection{Supersymmetric versions}

A possibility to have the EW scale  naturally smaller than $F_a$ without fine-tuning 
(and also $F_a\ll M_P$) is to 
supersymmetrize the above models. 
For the KSVZ models this implies  that the interactions of $\Phi$ with the quarks $Q'_{L,R}$ 
must arise from  the superpotential term (for $n=1$)
\be
W= y_{Q'} \Phi \bar Q'_L Q'_R\,,
\label{superpot1}
\ee
while for  DFSZ models 
\be
W= \kappa \Phi H_1H_2\,.
\label{superpot2}
\ee
Notice that in this latter case,  when $\Phi$ gets a VEV, \eqref{superpot2} generates a supersymmetric mass for the Higgs doublets. Since this mass must be of order the EW scale, this requires 
$\kappa \sim {\rm TeV}/F_a$, making this term irrelevant in the scalar potential.

The above superpotentials, however,  leave  the VEV of $\Phi$ undetermined.
The latter can be generated once we add  soft supersymmetry breaking (SSB) terms,
which are also required to get realistic models for the EW scale.
The relevant potential for $\phi$ is then simply given by\footnote{For the KSVZ model we must assume that  the SSB masses of $Q'_{L,R}$ are positive such that  colored scalars do not  get  VEVs.}
\be
V=\frac{1}{2}m^2_\phi(\phi)\, \phi^2\,,
\label{potSSB}
\ee
where $m^2_\phi(\phi)$  is the SSB mass of $\phi$ and its dependence on $\phi$ arises from loop effects.
The potential \eqref{potSSB} can lead to a nonzero minimum for $\phi$, 
similar to the Coleman-Weinberg model,
by demanding that 
$m^2_\phi$ is positive at large $\phi$ but "runs" towards negative values as $\phi$ decreases.
The VEV of $\phi$ then occurs at around  $m^2_\phi(\langle\phi\rangle)\sim 0$,  or, more precisely, at
\be
m^2_\phi(\langle\phi\rangle)=-\frac{1}{2}\beta_{m^2_\phi}(\langle\phi\rangle)\,,
\label{minphi}
\ee
where $\beta_{m^2_\phi}=dm^2_\phi/d\ln\phi$ arises at the  quantum level and it is then one-loop suppressed.
For example, from the interaction \eqref{superpot1}, we have
\be
\beta_{m^2_\phi}=\frac{3y_{Q'}^2}{8\pi^2}\left( m_{\widetilde Q'_L}^2+ m_{\widetilde Q'_R}^2+m^2_\phi+|A_{y_{Q'}}|^2\right)\,,
\ee
where $m_{\widetilde Q'_{L,R}}$ and $A_{y_{Q'}}$ are respectively the SSB mass of the scalar component of $Q'_{L,R}$ 
and the trilinear SSB term.
It is easy to choose the SSB parameters   such that the minimum of the potential \eqref{minphi}
occurs at the desired value $\langle\phi\rangle= F_a$.

Let us consider the  phase transition of this model.
At high temperatures the potential is given by
\be
V(T)=\left(D_\phi T^2+\frac{1}{2}m^2_\phi(\phi)\right)\phi^2+\cdots
\ee
where $D_\phi$ is defined in \eqref{thermlapot}.\footnote{We are neglecting cubic and quartic corrections which can be induced at the one-loop level by thermal corrections and supersymmetry breaking terms. 
These terms  will not change  our conclusions.} 
The critical temperature is at 
\be
T_c\simeq \sqrt{-m^2_{\phi, \rm min}/2D_\phi}\sim {\rm TeV}\,,
\ee
where $m^2_{\phi, \rm min}$ corresponds to the minimal value of $m^2(\phi)$.
As long as  this minimal value is negative and occurs  at $\phi>0$, as we will assume from now on,
 the potential at $T_c$ will have a thermal barrier,
and a first-order phase transition will be possible. 
We can estimate the bounce action of a thermal $O(3)$-symmetric bubble as~\cite{Anderson:1991zb}
\be
{S_B = \frac{S_3}{T}}\sim 4\pi \min_{\phi_{\rm tun}} \frac{|\phi_{\rm tun}|^3}{T\sqrt{|V(\phi_{\rm tun})|}}
\sim 4\pi\min_{\phi_{\rm tun}}\frac{\phi_{\rm tun}^2}{T\sqrt{|m^2_\phi(\phi_{\rm tun})|}}
\,,
\ee
where the minimization is over the tunneling point $\phi_{\rm tun}$. The latter
in this case corresponds to the smallest possible  $\phi_{\rm tun}$, determined by  $V(\phi_{\rm tun})\approx V(0)$:
\be
m^2_\phi (\phi_{\rm tun})\approx -2 D_\phi T^2\,.
\ee
Since we have assumed that $|m^2_\phi(\phi)|$ decreases with $\phi$ after it has reached $|m^2_{\phi, \rm min}|$,
$\phi_{\rm tun}$ also decreases as $T$ drops.
Therefore $S_B$ decreases till it reaches $S_{n}$ where bubbles form and complete the phase transition.
We can estimate the resulting value of $\alpha$ as
\be
\alpha\sim \frac{V(\langle\phi\rangle)}{T^4_c}\sim \frac{F_a^2}{\rm TeV^2}\gg 1\,,
\ee
and the value of $\beta/H_*$ as
\be
\frac{\beta}{H_*}\simeq \frac{4m^2_\phi (\phi_{\rm tun})}{ \beta_{m^2_\phi}(\phi_{\rm tun})}S_{n}\gg 1\,.
\label{betaSUSY}
\ee
From \eqref{Tstart}, we have
\be
T_*\simeq 
10^7\  {\rm GeV}\left(\frac{100}{g_*} \right)^{1/4}
\sqrt{\left(\frac{F_a}{10^{12}\, {\rm GeV}}\right)\left(\frac{m_\phi}{\rm TeV}\right)}\,,
\ee
which lies close to the LIGO and ET range for  interesting values of $F_a$.
Nevertheless, the predicted values of $\beta/H_*$ from \eqref{betaSUSY} are quite large, $\gtrsim 100$,
which makes it impossible to be seen at LIGO,  since bubble collisions would be the main source of GWs in this case, and only ET could be able to detect this type of phase transition -- see Fig.~\ref{sensitivities}

\subsection{Strongly-coupled PQ models}
\label{SCPQM}

After discussing the possibility of a first-order phase transition in the KSVZ and DFSZ models, let us now move to a different class of realizations of the PQ mechanism. We consider the case in which the PQ symmetry arises as an accidental global symmetry of a  new strong sector  that,
similarly to the $U(1)_A$ in QCD, is broken  at the scale where  condensates are formed. This scale can be chosen to be of order $F_a$. 

GWs can arise in this case from the  deconfined-to-confined phase transition which proceeds in the following way. At high temperatures  ($T\gg F_a$)   the strong sector is expected to be in a deconfined phase,
where the constituents are not confined into hadrons. 
As the temperature drops below $T_c \sim F_a$, the confined phase becomes energetically favorable,
and the model can go through a  phase transition.
For a gauge theory with a large number of colors $N$, this phase transition is expected
to  be of first order,  
and indeed this  can be proven to be the case  for holographic models \cite{Witten:1998zw,Creminelli:2001th,Konstandin:2011dr}.
To  address this phase transition quantitatively,
we will follow the strongly coupled models  studied in Ref.~\cite{vonHarling:2017yew,Baratella:2018pxi,Agashe:2019lhy} which have a  weakly-coupled five-dimensional version via holography (see \cite{Randall:2006py} for the GW signal arising from such a phase transition at the TeV scale).
This helps to  reduce the number of  parameters, although the conclusions can be extended to models without 
holographic versions~\cite{Baratella:2018pxi}.

The requirements for the strongly-coupled PQ model  are the following.
We assume that the strong sector  has  a  global  $U(1)_{PQ}\otimes SU(3)_c$  symmetry 
with an  $U(1)_{PQ}-SU(3)_c-SU(3)_c$ anomaly (this means that its constituents must be colored under $SU(3)_c$).
We also assume that 
the confinement scale $\Lambda_c$ of the new strongly-coupled sector is determined by a potential for the dilaton $\mu$ given by  
\be
V_{\rm eff}(\mu)=\frac{N^2}{16\pi^2} \lambda(\mu)\mu^4\,,
\label{potdilaton}
\ee
where the dependence of the quartic coupling $ \lambda(\mu)$  on $\mu$ is dictated by the explicit breaking of scale invariance (several examples are given in  \cite{Baratella:2018pxi}).
  We identify  the mass gap $\Lambda_c$ with the  dilaton VEV, 
$\langle \mu\rangle=\Lambda_c$.
We further assume that confinement also leads to the spontaneous breaking of $U(1)_{PQ}$. The axion is then the corresponding (composite meson) Nambu-Goldstone boson.\footnote{Holographic versions of these models can be found in~\cite{Choi:2003wr,Flacke:2006ad,Kawasaki:2015lea,Bigazzi:2019eks,Cox:2019rro}.}
The $U(1)_{PQ}-SU(3)_c-SU(3)_c$ anomaly  guarantees  the coupling \eqref{definition}, with
an axion decay constant 
\be
F_a=\frac{\sqrt{N}}{4\pi}\Lambda_c\,,
\label{defFa}
\ee
where $N\gg 1$ plays the role of the   number of "colors" of the strong sector.

\begin{figure}[t]
\centering
\includegraphics[width=10cm]{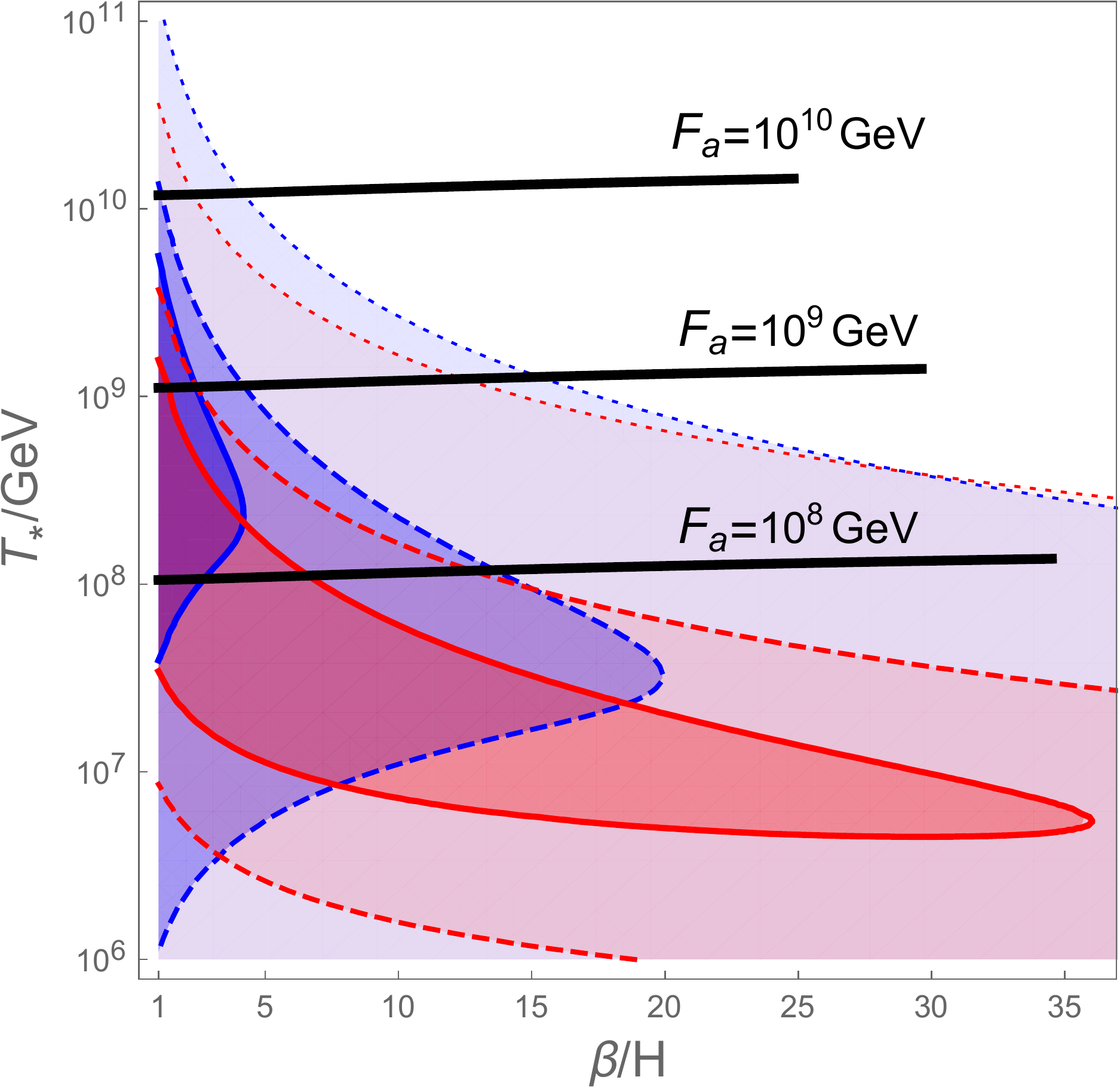}
\caption{\it 
Predicted values of $T_*$ and $\beta/H_*$ (black lines) for  strongly-coupled PQ models (see Section \ref{SCPQM} for details)   for $F_a=10^8$, $10^9$ and $10^{10}$  {\rm GeV}.
The  present and expected future experimental GW reaches are depicted as red and blue areas, using the same color code as in Fig.~\ref{sensitivities}.}
\label{strongPQGW}
\end{figure}

The free-energy of the unconfined phase is given by ${\cal F}_{\rm dec}\simeq-\pi^2N^2T^4/8$,
while in the confined phase ${\cal F}_{\rm conf}=V_{\rm eff}(\langle\mu\rangle)$.
Thus, the critical temperature at which the confined phase
is energetically favorable  follows as \cite{Baratella:2018pxi}
\be
T_c\simeq 0.3\times 10^{10} \,{\rm GeV}\left(\frac{\left(\Lambda_c\, m_{\rm dil}\right)^{1/2}}{10^{10}\ {\rm GeV}}\right)\,,
\label{tcrit}
\ee
where $m_{\rm dil}$ is the dilaton mass.
The rate of the phase transition from the unconfined to the confined phase 
is in most of the cases dominated by vacuum tunneling
whose bounce action is roughly given by \cite{Baratella:2018pxi}
\begin{equation}\label{twdilaton}
S_B\sim   \frac{24 N^2}{|{\lambda}(\mu_{\rm tun})|}\,,
\end{equation}
where $\mu_{\rm tun}\simeq T{\Lambda_c}/{T_c}$.
We are interested in phase transitions  with large values of $\alpha$ and small values of $\beta/H_*$, as 
this maximizes  the GW strength.
As in the case studied in Sec.~\ref{sub:cw}, this arises  when there is a  period of supercooling, 
which in this case happens when the universe stays for a while in the unconfined phase before the phase transition takes place.
In order to achieve that, $|{\lambda}(T)|$ must slowly increase  as  $T$ decreases, so that
$S_B$ slowly approaches $S_{n}$.
In this case we have $\alpha\gtrsim 1$ while 
\be
\frac{\beta}{H_*}\simeq \frac{\beta_\lambda(T_n)}{\lambda(T_n)}S_{n}-4\,,
\label{betaH}
\ee
where $\beta_\lambda=d\lambda/d\ln\mu$. 
From this, we see that  long periods of supercooling, where  $S_B$  evolves slowly towards
$S_{n}$, can give rise to small values of  $\beta/H_*$.
This can be appreciated in Fig.~\ref{strongPQGW},  where we consider $\lambda(\mu)=b_0(\ln(\Lambda_c/\mu)-1/4)$
and vary  $b_0$, or equivalently,  $T_n$.   
Starting at  $T_n= 0.02\, \Lambda_c$ and going to  smaller values, 
we move from the right to the left along the black solid lines of Fig.~\ref{strongPQGW} 
(taking $N=3$ and choosing different values of $F_a$).\footnote{The model works for moderately large values of $N$,  since $N$ must  be large enough in order for  the holographic model to be perturbative, but  not too large, otherwise the bounce action \eqref{twdilaton} becomes too large and the universe gets trapped forever in the unconfined phase. See \cite{Baratella:2018pxi} for details.} 
The value of  $T_*$ is the  reheating temperature after the phase transition is completed which is found to be 
$T_*\simeq 1.8 {\sqrt{N}}/{g_{*}^{1/4}}T_{c}$ \cite{Baratella:2018pxi}.
Using  this and \eqref{defFa}, we obtain the relation
\be
T_*\simeq 2\, F_a
\left(\frac{100}{g_*} \right)^{1/4}
\left(\frac{m_{\rm dil}}{\Lambda_c}\right)^{1/2}\,.
\ee

Even though this scenario realizes supercooling, which strongly dilutes the thermal plasma around the bubbles and leads to $v_{w}\simeq 1$, it is possible that sound waves and turbulence are still the main source of GWs. This is important because in this case detection could be easier, as can be appreciated in Fig.~\ref{sensitivities}.
The reason for this is that the deconfined-to-confined phase transition
 involves gauge bosons (dark gluons) which receive a mass across the bubble walls. As pointed out in~\cite{Bodeker:2017cim}, these can be radiated off as particles cross the bubble walls. This so-called transition radiation generates friction on the motion of the bubble walls and can halt their acceleration. More concretely, transition radiation leads to an upper bound on the $\gamma$ factor of the bubble walls, given by~\cite{Baratella:2018pxi}
\begin{equation}
\label{eq:gammac}
\gamma_{c}\sim \left(\frac{\Lambda_{c}}{T_{n}}\right)^3\,.
\end{equation}
If bubbles collide significantly after reaching $\gamma_c$, then most of the energy available in the phase transition goes to the thermal plasma, since the bubbles are not in the runaway regime even if $v_{w}$ is very close to one. However, bubbles can also collide before they have time to reach $\gamma_{c}$. In this case, bubble collisions are the dominant source of GWs. Let us then estimate the amount of supercooling required to be in this latter regime. Following~\cite{Baratella:2018pxi}, the maximal $\gamma$ factor achieved before collision is
\begin{equation}
\label{eq:gammamax}
\gamma_{\text{max}}\sim \left(\frac{H_{*}}{\beta}\right)\frac{M_{P}}{\Lambda_{c}}\frac{T_{n}}{\Lambda_c}\,.
\end{equation}
Matching the equation above to \eqref{eq:gammac} we obtain
\begin{equation}
\label{Tngammamax}
T_{n, \gamma_{c}=\gamma_{\text{max}}}\sim \Lambda_{c}\left(\frac{\beta}{H_{*}}\right)^{1/4}\left(\frac{\Lambda_{c}}{M_{P}}\right)^{1/4}\sim F_{a}\left(\frac{\beta}{H_{*}}\right)^{1/4}\left(\frac{F_{a}}{M_{P}}\right)^{1/4}\,.
\end{equation}
Thus we see that for $F_{a}\sim 10^{8}-10^{10}$ GeV, sound waves and turbulent motion in the plasma are expected to be the dominant source of GWs when $T_{n}\gtrsim 10^{-2}-10^{-3}\, F_{a}$. For longer supercooling, bubble collisions are the main source instead.
}

In Fig.~\ref{strongPQGW}, we show the predictions of $T_*$ vs.~$\beta/H$ for the strongly-coupled PQ models as well as the present and expected future sensitivities from GW searches. Solid lines are for current LIGO, dashed ones for LIGO at design sensitivity and dotted ones for ET. The corresponding regions in blue can be probed if the GW signal is mainly generated from bubble collisions, while those in red can be tested if GW production is dominated by sound waves. We have assumed the GW spectra from these sources as summarized in Sec.~\ref{sensi}. As we have discussed, sound waves can be the main source of GWs even in the supercooled regime. In this case, however, the amplitude of the resulting GWs may be suppressed compared to the one given in \eqref{OmegaSW} \cite{Caprini:2019egz,Ellis:2018mja,Ellis:2019oqb}. We therefore note that the sensitivity regions for  sound-wave production of GWs shown in Fig.~\ref{strongPQGW} are only an upper bound. They may turn out to be somewhat smaller once sound-wave production of GWs in this regime is better understood. In the very supercooled regime where  $T_n \lesssim T_{n, \gamma_{c}=\gamma_{\text{max}}}$, on the other hand, bubble collisions are the dominant source of GWs which we expect to be well described by \eqref{OmegaPhi}. The corresponding sensitivity regions in Fig.~\ref{strongPQGW} are therefore more robust. 
We see from Fig.~\ref{strongPQGW} that the  phase transition of the strongly-coupled PQ models can be detected by LIGO (at current and design sensitivity) if there is enough supercooling. The smaller $F_{a}$ is, the more likely is the  detection of the GWs.

Finally, let us conclude this subsection by noting that in principle 
an alternative option for a long period of supercooling is to have $\lambda(T)$ evolving too slow
(for a holographic example see \cite{Pomarol:2019aae}))
such that the condition $\Gamma \simeq H^4$ is not met and the universe gets trapped in the unconfined phase.
As discussed in \cite{vonHarling:2017yew,Baratella:2018pxi}, the universe could still exit supercooling
at the QCD scale, where a new contribution to the dilaton potential arises.
In order for this to happen, we need the strong sector to have  
constituents which are charged under $SU(3)_c$.
This is indeed  the case for the axion models discussed here, since, as we have mentioned, 
the strong sector must have an $SU(3)_c$ symmetry in order for the axion to couple to $G\tilde G$.
Nevertheless, exit due to QCD effects is not possible here since $F_a$ is much larger than the scale 
where QCD becomes strong, and to exit supercooling at such low temperatures, $S_{B}$ would need to be 
of order one.

\section{Conclusions}
\label{conclu}

We have shown that LIGO has the possibility to detect GWs arising from a phase transition which occurs  in the early universe at temperatures around $10^{8}$ GeV.
As shown in Fig.~\ref{Talpha}, however, detection requires the  phase transition to be  strong enough
with values of  the latent heat parameter $\alpha>1$. 
For these types of phase transitions 
LIGO will be  able to  detect GWs  for  values of the inverse  transition time $\beta/H_*$ up to $\sim 10^3$.
On the other hand, the proposed ET observatory will be able to access  phase transitions with  slightly smaller values of $\alpha$
but much larger  $\beta/H_*$.  
In particular, as shown in Fig.~\ref{sensitivities},
ET will access phase transitions with $\alpha\gtrsim 0.1$,  and  $\beta/H_*\lesssim 10^6$.

The breaking of the PQ symmetry, required in QCD axion models, is a particularly well motivated example of such a phase transition. Indeed, the PQ phase transition would have to occur at temperatures $T\sim 10^8-10^{12}$ GeV, if the initial axion misalignment is not tuned to small values. The main message of this work is that LIGO, at current and design sensitivity, will be able to probe some of the simplest realizations of the PQ mechanism. 

In particular, we have shown that DFSZ realizations have the right ingredients to generate a GW signal, which is in the reach of LIGO. This occurs when the PQ symmetry breaking is of Coleman-Weinberg type, that is when the mass parameters of the model are small and the minimum is generated by quantum effects. Our key results are presented in Fig.~\ref{CWcase}, which shows that PQ scales up to $F_a\lesssim 10^{11}~\text{GeV}$ can be probed by LIGO and even more by ET. We note though that for this case some tuning may be required to obtain the needed small mass parameters.

Furthermore, we have discussed an alternative type of phase transition in the DFSZ model, which is due to a zero-temperature tree-level barrier. This would proceed via an intermediate step where the EW symmetry is broken at high scales, before tunneling from this phase to the PQ broken phase. We have shown that this case can exhibit $\alpha\gtrsim 1$, while the typical values of $\beta/H_{*}$ make its GW signal suited for detection at ET. A more detailed investigation of the parameter space which allows for a detectable two-step PQ transition is left for future work, as are also the phenomenological implications of the associated high-scale breaking of the EW symmetry.

For KSVZ realizations, we have shown that the simplest model does not lead to a strong first-order phase transition. However, supersymmetric KSVZ and DFSZ models can exhibit a first-order phase transition, with naturally small mass scales. We have found that the PQ symmetry breaking can be driven by  supersymmetry-breaking effects, giving a first-order phase transition with $\alpha\gg 1$ and $\beta/H_*\gtrsim100$.

We have continued our exploration of PQ phase transitions by considering models where the symmetry is broken by strong dynamics. In this case supercooling arises rather generically, without the need to tune mass parameters. The transition from the unconfined to the confined phase in these realizations can be strong enough to give a GW signal detectable at LIGO. Our key results for this type of phase transition are presented in Fig.~\ref{strongPQGW}.

Interestingly, other proposed observatories, like DECIGO~\cite{Kawamura:2006up} and BBO~\cite{Yagi:2011wg}, would be able to probe the small frequency tails of the broad GW spectra generated by the strongest first-order phase transitions which we have discussed in this work. Looking further into the future, GW detectors with sensitivity at higher frequencies than LIGO and ET, such as~\cite{Evans:2016mbw}, will open the possibility to discover phase transitions from QCD axion models with $F_{a}$ up to $10^{11}~\text{GeV}$ and weaker than the ones that we considered here.

Overall, as laboratory experiments progress in their search for the QCD axion at low energies, we have shown that LIGO can 
{\em already} join this effort by hearing the axion `birth' at the very high PQ scale.

{\it Note added:} While preparing this manuscript we became aware of the work of \cite{DelleRose:2019pgi} which also considers
 models with a PQ phase transition detectable at LIGO.

\medskip 
\section*{Acknowledgments}

We are grateful to Ken Olum for help with \texttt{AnyBubble}. 
We also would like to thank Francesc Ferrer  and Giuliano Panico for discussions on related work.
AP  was supported   by the Catalan ICREA Academia Program.
This work was also partly supported by the grants FPA2017-88915-P, 2017-SGR-1069 and 
SEV-2016-0588.

\appendix

\section{The scalar potential}
\label{app1}

In this Appendix we provide formulae to calculate the loop-corrected potential for scalar fields at finite temperature (see e.g. \cite{Quiros:1994dr} for a review and~\cite{ Curtin:2016urg} for a recent discussion).

Let us  consider a set of scalar fields $\{\phi_{i}\}$, with tree-level zero-temperature potential given by $V_{0}(\{\phi_{i}\})$. These scalar fields may or may not be coupled to extra fermionic and/or bosonic degrees of freedom. We keep the discussion general and number all the fields (the non-scalars coupled to the scalars as well as the scalars themselves) with an index $a$. The number of degrees of freedom associated with each field is $g_{a}$. Of particular importance for phase transitions is the dependence of the field masses on the values of the scalar fields $\{\phi_{i}\}$, which is usually of the form $m_{a}^{2}\sim c+b\phi_{i}^{2}$, with $c$ and $b$ constants. For the scalar fields, the masses $m_{a}^{2}$ are to be taken in the mass eigenstate basis, i.e.~they are the eigenvalues of the $i\times i$-dimensional mass matrix obtained from the tree-level scalar potential.

The tree-level zero-temperature potential receives the following corrections:
\begin{itemize}
\item[$\mathbf{1.}$]{\textbf{Coleman-Weinberg}: at zero temperature, the one-loop correction to $V_{0}(\{\phi_{i}\})$ using dimensional regularization and the $\overline{\text{MS}}$ renormalization scheme is given by:
\begin{equation}
\label{eq:cw}
V_{\text{CW}}\left(\{\phi_{i}\}\right)=\sum_{a}(-1)^{F}g_{a}\frac{m_{a}^{4}\left(\{\phi_{i}\}\right)}{64\pi^{2}}\left[\ln\left(\frac{m^{2}_{a}\left(\{\phi_{i}\}\right)}{\Lambda^{2}}\right)-c_{a}\right].
\end{equation}
Here $F=1$ for fermions and $F=0$ for bosons. Similarly, $c_{a}=3/2$ for scalars and fermions and $c_{a}=5/2$ for vectors.}
\item[$\mathbf{2.}$]{\textbf{Thermal}: at finite temperature $T$, the one-loop thermal correction to $V_{0}(\{\phi_{i}\})$ is given by:
\begin{equation}
\label{eq:VT}
V_{T}\left(\{\phi_{i}\},T\right)=\sum_{a}(-1)^{F}g_{a}\frac{T^{4}}{2\pi^{2}}J_{B/F}\left[\frac{m_{a}^{2}(\{\phi_{i}\})}{T^{2}}\right].
\end{equation}
Here the functions $J_{B/F}$ are defined as
\begin{equation}
\label{eq:jbf}
J_{B/F}(y^{2})=\int_{0}^{\infty}dx~x^{2}\ln\left[1\mp e^{-\sqrt{x^{2}+y^{2}}} \right].
\end{equation}
For certain purposes, it is enough to consider the following expansion of these functions in $m^{2}_{a}/T^{2}$:
\begin{align}
\label{eq:jbhight}
J_{B}(m^{2}/T^{2})&=-\frac{\pi^{4}}{45}+\frac{\pi^{2}}{12}\left(\frac{m}{T}\right)^{2}-\frac{\pi}{6}\left(\frac{m^{2}}{T^{2}}\right)^{3/2}-\frac{1}{32}\left(\frac{m}{T}\right)^{4}\ln\left(\frac{m^{2}}{a_{b}T^{2}}\right)+\dots\,,\\
\label{eq:jfhight}
J_{F}(m^{2}/T^{2})&=\frac{7\pi^{4}}{360}-\frac{\pi^{2}}{24}\left(\frac{m}{T}\right)^{2}-\frac{1}{32}\left(\frac{m}{T}\right)^{4}\ln\left(\frac{m^{2}}{a_{f}T^{2}}\right)+\dots\,,
\end{align}
where $\ln(a_{b})=5.4076$ and $\ln(a_{f})=2.6351$.}
\end{itemize}

Eqs.~\eqref{eq:jbhight} and \eqref{eq:jfhight} deliver an important message for phase transitions driven by thermal corrections: since $m_{a}^{2}\sim c+b\phi^{2}$, the leading thermal corrections due to bosons take the form 
\be
V_{T}= D_{\phi}T^{2}\, \phi^{2}+E_\phi\,T\phi^{3}+\cdots\,.
\label{thermlapot}
\ee
Both fermions and bosons can contribute to $D_\phi$. 
On the other hand, only bosons can contribute to $E_\phi$ and induce a cubic term in $\phi$. 
This latter term is important, since it can induce a barrier separating two minima in field space.
A further more subtle point is related to the infrared singularity in the high temperature limit of $V_{T}$~\cite{Gross:1980br, Parwani:1991gq, Arnold:1992rz}, as defined in \eqref{eq:VT}. The standard strategy to avoid this problem is to replace the bosonic squared masses $m^2_{i}$ with the dressed squared masses $m_{i}^{2}(\phi_{j})+2 D_{\phi_i} T^2$ (before diagonalization of the scalar mass matrix), where $D_{\phi_i}=2 [\partial_{\phi_i}^{2} V_{T}/T^2]_{\phi_{i}, T=0}$. This replacement is done everywhere in $V_{T}$ as well as in $V_{\text{CW}}$. These so-called daisy corrections generically weaken the strength of a phase transition, since at high temperatures $T^{2}\gtrsim m_{a}^{2}$, they screen the field dependence of the leading order cubic terms in the bosonic thermal potential.

For reference, let us conclude this section by providing the expressions for the daisy masses of the real, $U(1)_{\rm EM}$-neutral components of $\Phi$, $H_1$ and $H_2$ which we have used in our work (we do not list the daisy masses of the imaginary and charged components, while those of the EW gauge bosons can be found in \cite{Quiros:1994dr}):
\begin{align}
D_{\phi}&=\frac{\kappa_{1}+\kappa_{2}}{12}+\frac{\lambda_{\phi}}{6}\,,\label{dphi}\\
D_{h_{1}}&=\frac{1}{96}\left(9g^{2}+3g'^{2}+\frac{12\lambda_{t}^{2}}{\cos^{2}\theta}+24\lambda_{1}+4\kappa_{1}+8\lambda_{12}\right)\,,\label{dh1}\\
D_{h_{2}}&=\frac{1}{96}\left(9g^{2}+3g'^{2}+24\lambda_{2}+4\kappa_{2}+8\lambda_{12}\right)\,.
\end{align}

\providecommand{\href}[2]{#2}\begingroup\raggedright\endgroup

\end{document}